\documentclass[a4paper,12pt]{article}

\usepackage{amsmath,amssymb}
\usepackage{graphicx}

\DeclareMathOperator{\Ai}{Ai}
\DeclareMathOperator{\B}{B}
\DeclareMathOperator{\lcm}{lcm}

\DeclareMathOperator{\Res}{Res}
\DeclareMathOperator{\Tr}{Tr}

\numberwithin{equation}{section}
\allowdisplaybreaks

\usepackage{geometry}
\newcounter{aff}

\begin{document}
\begin{titlepage}
\begin{flushright}
{\footnotesize YITP-15-105}
\end{flushright}
\begin{center}
{\Large\bf Instanton effects in ABJM theory with general $R$-charge assignments
}

\bigskip\bigskip
{\large 
Tomoki Nosaka\footnote[1]{\tt nosaka@yukawa.kyoto-u.ac.jp}
}\\
\bigskip
{\small\it Yukawa Institute for Theoretical Physics,
Kyoto University,\\
Kyoto 606-8502, Japan}
\end{center}

\begin{abstract}

We study the large $N$ expansion of the partition function of the quiver superconformal Chern-Simons theories deformed by two continuous parameters which correspond to the general $R$-charge assignment to the matter fields.
Though the deformation breaks the conformal symmetry, we find that the partition function shares various structures with the superconformal cases, such as the Airy function expression of the perturbative expansion in $1/N$ with the overall constant $A(k)$ related to the constant map in the ABJM case through a simple rescaling of $k$.
We also identify five kinds of the non-perturbative effects in $1/N$ which correspond to the membrane instantons.
The instanton exponents and the singular structure of the coefficients depend on the continuous deformation parameters, in contrast to the superconformal case where all the parameters are integers associated with the orbifold action on the moduli space.
This implies that the singularity of the instanton effects would be observable also in the gravity side.

\end{abstract}

\bigskip\bigskip\bigskip

\end{titlepage}
\tableofcontents

\section{Introduction}
Though the fundamental principles of M2-brane interactions are not clear, a particular class of $U(N)$ superconformal quiver Chern-Simons theories are proposed as the worldvolume theory of the $N$-stack of interacting M2-branes.
One of the supporting evidences is that the large $N$ limit of the free energy computed in these theories exhibit the $N^{3/2}$ scaling.
This precisely reproduces the result obtained in the eleven dimensional supergravity on $\mathrm{AdS}_4\times Y_7$ \cite{DMP1,HKPT,MS} (or their consistent truncations in 4d), where $Y_7$ is some seven dimensional manifold associated with the theory.
Taking this gauge/gravity correspondence inversely, the field theory analysis beyond the large $N$ limit is expected to shed new lights on the M-theory beyond the classical supergravity.

Among the theories of $N$ M2-branes the ABJM theory \cite{ABJM} is the most symmetric one, and hence have been studied with the greatest efforts.
The ABJM theory is the ${\cal N}=6$ $U(N)_k\times U(N)_{-k}$ quiver superconformal Chern-Simons theory.
In the ${\cal N}=2$ notation, each vertex of the quiver is assigned with $U(N)$ Chern-Simons vector multiplet $(A_\mu,\sigma,\lambda,D)$ with Chern-Simons levels $\pm k$ while each edge is assigned with a pair of bifundamental hypermultiplets $(\phi,\psi,F)$ and $({\widetilde \phi}, {\widetilde\psi}, {\widetilde F})$ which are charged under $U(1)_R$ as $(R_{\phi,{\widetilde \phi}^\dagger}, R_{\psi,{\widetilde \psi}^\dagger}, R_{F,{\widetilde F}^\dagger})=(1/2,-1/2,-3/2)$ \cite{ABJM,HLLLP}.
The dual geometry to this theory is $\mathrm{AdS}_4\times S^7/\mathbb{Z}_k$.
With the help of the localization technique, the partition function of the ABJM theory can be reduced to a matrix model with $2N$ integration variables \cite{KWY1}.
After the determination of the leading $N^{3/2}$ behavior \cite{DMP1}, the matrix model was further analyzed in the 't Hooft limit $k,N\rightarrow\infty$ with $\lambda=N/k$ fixed \cite{DMP1,DMP2,FHM}, with the help of the relation between the 't Hooft expansion of the matrix model and the free energy of the topological string theory on the local $\mathbb{P}^1\times \mathbb{P}^1$ \cite{MP1}.

Later a new expression of the ABJM matrix model was discovered as the canonical partition function of a quantum statistical system of $N$ particle ideal Fermi gas, where the level $k$ is converted into the Planck constant $\hbar=2\pi k$ in the statistical system \cite{MP2}.
This relation enables us a systematic analysis of the large $N$ expansion of the partition function in the M-theoretical regime $k<\infty$, in terms of the grand potential $J(\mu)$ defined by
\begin{align}
e^{J(\mu)}=1+\sum_{N\ge 1}e^{\mu N}Z(N).
\label{ZtoJ}
\end{align}
Here $\mu$ is an auxiliary parameter called the chemical potential dual to $N$.
The original partition function can be recovered by the following inverse transformation
\begin{align}
Z(N)=\int \frac{d\mu}{2\pi i}e^{J(\mu)-\mu N}.
\label{JtoZ}
\end{align}
For finite value of $k$, the large $N$ expansion of the partition function corresponds to the large $\mu$ expansion of the grand potential. 

After various efforts \cite{MP2,HMO1,PY,HMO2,CM,HMO3}, finally all the $1/\mu$ corrections were completely determined \cite{HMMO}, including both the perturbative and non-perturbative effects.
The perturbative part of the grand potential is a cubic polynomial in $\mu$
\begin{align}
J^\text{pert}(\mu)=\frac{C}{3}\mu^3+B\mu+A,
\label{Jpert}
\end{align}
with $C$, $B$ and $A$ some constants.
In the partition function this turns into the all order perturbative sum expressed as an Airy function (as obtained in \cite{FHM})
\begin{align}
Z^\text{pert}(N)=e^AC^{-\frac{1}{3}}\Ai[C^{-\frac{1}{3}}(N-B)].
\label{Airy}
\end{align}

There are two kinds of non-perturbative effects in the grand potential: $e^{-4m\mu/k}$ and $e^{-2n\mu}$ $(m,n=1,2,\cdots )$.
Through the inversion formula, these effects turn to the corrections of ${\cal O}(e^{-\sqrt{N/k}})$ or ${\cal O}(e^{-\sqrt{kN}})$ in the partition function.
In the gravity side, the non-perturbative effects are quantitatively interpreted as the effects of the fundamental M2-branes winding on $Y_7$.
Indeed, the exponents of the non-perturbative effects in the partition function are proportional to $R_\text{AdS}^3$ and hence can be explained in terms of the excitation energy of the winding M2-branes.
The first kind of the non-perturbative effects ${\cal O}(e^{-4\mu/k})$ correspond to the M2-branes winding the $\mathbb{Z}_k$-orbifolded cycle and thus called the worldsheet instanton effects \cite{CSW}, while the second ones ${\cal O}(e^{-2\mu})$ correspond to the M2-branes winding in other three directions and called the membrane (D2) instanton effects \cite{BBS,DMP2}.
Although the Chern-Simons level $k$ is originally integer, in the ABJM matrix model we can generalize $k$ to be an irrational number.
This allows the separative analysis of the two kinds of non-perturbative effects, respectively by the 't Hooft expansion of the partition function and the semiclassical expansion of the grand  potential.
For the complete determination of the coefficients in front of these exponentials, however, it was essential to observe the following singular structures of them at finite and integral $k$ \cite{HMO2,CM}.
For integral $k$, the exponents of two kind of non-perturbative effects coincide when $2m=kn$.
In this case, the individual coefficients are divergent, while the divergence are completely cancelled between the two coefficients.
This structure, called as the HMO cancellation mechanism in \cite{HMMO}, was used in the extrapolation of the small $k$ expansion of the coefficient of the second kind of instantons for higher $n$ and to conjecture their uniformed expression.

Recently similar structures in the large $N$ expansion were discovered in the more general superconformal quiver Chern-Simons theories.
The Airy function expression of the all order perturbative corrections in $1/N$ was already claimed for the general $U(N)$ ${\cal N}=3$ circular quiver superconformal Chern-Simons theories in \cite{MP2}.
The non-perturbative effects were analyzed in detail for a special class of ${\cal N}=4$ superconformal quiver Chern-Simons theory \cite{HM,HaOk1,MN1,MN2,MN3,HHO}.\footnote{
The Airy function structure and the instanton effects are also revealed for the cases of non-circular quivers \cite{ADF,MN4} or non-unitary gauge groups \cite{MoSu,O}.
}
Each of these theories are characterized by a integer $k$ and the signs $s_a=\pm 1$ assigned on the edges with which the Chern-Simons level on the $a$-th vertex is given as \cite{IK2}
\begin{align}
k_a=k(s_a-s_{a-1})/2.
\label{kins}
\end{align}
A set of signs $\{s_a\}_{a=1}^M$ is labelled by positive integers $m$, $\{q_a\}_{a=1}^m$ and $\{p_a\}_{a=1}^m$ as\footnote{
For $k=1$, $m=1$ and $(q_1,p_1)=(N_f,1)$ the matrix model is identical with the $N_f$ matrix model studied in \cite{GM,HaOk1}.
}
\begin{align}
\{s_a\}_{a=1}^M&=\{
\underbrace{1,1,\cdots,1}_{q_1},
\underbrace{-1,-1,\cdots,-1}_{p_1},
\underbrace{1,1,\cdots,1}_{q_2},
\underbrace{-1,-1,\cdots,-1}_{p_2},\nonumber \\
&\quad\quad\cdots,\underbrace{1,1,\cdots,1}_{q_m},
\underbrace{-1,-1,\cdots,-1}_{p_m}
\}
\end{align}
which we shall abbreviate as
\begin{align}
\{s_a\}_{a=1}^M=\{(+1)^{q_1},(-1)^{p_1},(+1)^{q_2},(-1)^{p_2},\cdots,(+1)^{q_m},(-1)^{p_m}\}.
\label{sas}
\end{align}
The dual geometry of this theory is the product spacetime of $\mathrm{AdS}_4$ and the radial section of $(\mathbb{C}^2/\mathbb{Z}_q\times \mathbb{C}^2/\mathbb{Z}_p)/\mathbb{Z}_k$, which was determined by analyzing the moduli space or the brane construction \cite{IK}.
Here $q$ and $p$ are the number of edges with $s_a=\pm 1$
\begin{align}
q=\sum_{a=1}^mq_a,\quad
p=\sum_{a=1}^mp_a.
\end{align}
The instantons effects in these theories were found to subdivide into four kinds $e^{-2\mu/q}$, $e^{-2\mu/p}$, $e^{-\mu}$ \cite{MN2} and $e^{-4\mu/(kqp)}$ \cite{HHO} and have richer divergent structures than in the ABJM case, controlled by $(k,q,p)$.

So far such detailed analyses, especially of the instanton effects, were successful only in the superconformal quiver Chern-Simons theories.
On the other hand, it was known that the leading $N^{3/2}$ scaling behavior of the free energy is satisfied even in some theories without conformal invariance.
Such theories are expected to be dual to the geometries which are asymptotically $\text{AdS}_4$ while have non-trivial structure in the bulk and exhibit completely different asymptotics in the opposite limit in the radial direction.
Therefore it is non-trivial and would be interesting whether the above structures hold, or how they are generalized, in such non-conformal theories.

In this paper we consider the following continuous deformation of the theory.
Starting from the ${\cal N}=4$ circular quiver superconformal Chern-Simons theory with the levels \eqref{kins}, we modify the $R$-charge assignments on the bifundamental hypermultiplets $(\phi_a,\psi_a,F_a)$ and $({\widetilde \phi}_a,{\widetilde \psi}_a,{\widetilde F}_a)$ on $a$-th edge as
\begin{align}
(R_{\phi_a},R_{\psi_a},R_{F_a})&=\biggl(\frac{1+\zeta_a}{2},\frac{-1+\zeta_a}{2},\frac{-3+\zeta_a}{2}\biggr),\nonumber \\
(R_{\widetilde \phi_a},R_{{\widetilde\psi}_a},R_{\widetilde F_a})&=\biggl(\frac{-1+\zeta_a}{2},\frac{1+\zeta_a}{2},\frac{3+\zeta_a}{2}\biggr)
\label{Rgen}
\end{align}
with
\begin{align}
-1<\zeta_a<1.
\label{zetabound}
\end{align}
In the flat space these are just a matter of convention, for which we shall call the $U(1)_R$ symmetry among the $U(1)$ global symmetries of the theory.
Once we realize the theory on a three sphere, however, the choice is relevant to the curvature couplings and results in a distinctive theory for each choice.
The theory is conformal only for $\zeta_a=0$, the canonical choice of the $R$-charges.
The partition function of this theory has been studied in detail in the limit of $N\rightarrow\infty$ in the context of the $F$-theorem \cite{J,JKPS}, and the leading $N^{3/2}$ scaling was obtained with the explicit expression of its coefficient \cite{GHP}.

To analyze the large $N$ expansion of the partition function, we first provide the Fermi gas formalism of this theory, with which we can compute the large $N$ expansion of the partition function systematically through the grand potential $J(\mu)$.
Restricting ourselves to the minimal separation of $s_a=\pm 1$
\begin{align}
\{s_a\}_{a=1}^M=\{(+1)^q,(-1)^p\},
\label{sminimal}
\end{align}
we find that the perturbative corrections again sum up to an Airy function \eqref{Airy}, with the three parameters $A$, $B$ and $C$ given by
\begin{align}
C&=\frac{2qp}{\pi^2k(q^2-\xi^2)(p^2-\eta^2)},\quad
B=\frac{\pi^2C}{3}-\frac{qp}{6k}\biggl(\frac{1}{q^2-\xi^2}+\frac{1}{p^2-\eta^2}\biggr)+\frac{kqp}{24},\nonumber \\
A&=\frac{p^2}{4}\Bigl(A_\text{ABJM}((q+\xi)k)+A_\text{ABJM}((q-\xi)k)\Bigr)+\frac{q^2}{4}\Bigl(A_\text{ABJM}((p+\eta)k)+A_\text{ABJM}((p-\eta)k)\Bigr).
\label{CBA}
\end{align}
Here $\xi$ and $\eta$ are the parameters associated with the total deformation over the edges with $s_a=\pm 1$ respectively as
\begin{align}
\xi=-\sum_{a=1}^{q}\zeta_a,\quad
\eta=\sum_{a=q+1}^{q+p}\zeta_a,
\label{xieta}
\end{align}
and $A_\text{ABJM}(k)$ is the quantity called the constant map in the ABJM theory \cite{HHHNSY}.
These expressions for the coefficients are natural generalizations of the results in the conformal case \cite{MN1}.

Analyzing the small $k$ expansion exactly in $\mu$, we also discover five kinds of the non-perturbative effects $e^{-2\mu/(q\pm\xi)}$, $e^{-2\mu/(p\pm\eta)}$ and $e^{-\mu}$ which are the generalization of the membrane instantons in the ABJM theory.
The instanton exponents \eqref{instcoefs} depends on $\xi$ and $\eta$, and the individual coefficient diverges at some special values of $\xi$ and $\eta$ as in the superconformal theories.

On the other hand, the counterparts for the worldsheet instantons are invisible in the small $k$ expansion, as they are non-perturbative in $k$.
These effects will be accessible from the exact values of the partition functions with various finite $(k,N)$.
We generalize the method for the systematic computation of these values known in the superconformal case \cite{TW,PY,MN3} to the general choice of $R$-charges \eqref{Rgen}.
The result is consistent with the Airy function and strongly support the conjectural expression for $A$ \eqref{CBA}.
On the other hand, the deviations from the Airy function are significantly different from the non-perturbative corrections obtained in the small $k$ expansion and will correspond to the worldsheet instantons.

The remaining part of this paper is organized as follows.
In the next section we introduce the Fermi gas formalism for the non-canonical $R$-charge assignments \eqref{Rgen}, in slightly more general framework of the ${\cal N}=3$ $U(N)$ circular quiver superconformal Chern-Simons theory.
In the subsequent sections, we concentrate on the theory of the minimal separations \eqref{sminimal} and compute the exact large $N$ expansion of the partition function using the Fermi gas formalism.
In section \ref{Fermisurface} we compute the perturbative corrections in $1/\mu$ and obtain the Airy function expression \eqref{Airy} with the explicit expression of the coefficients $B$ and $C$ in \eqref{CBA}.
In section \ref{WKB} we analyze the small $k$ expansion of the grand potential in more detail and conjecture the expression of $A$ in \eqref{CBA}.
We also determine the explicit coefficients of the five kinds of the membrane instantons and argue the mixing and divergent structures of these instantons.
In section \ref{finitek} we explain the method of the exact computation of the partition function and compare the results with the small $k$ expansion.
Finally in section \ref{discuss} we summarize our results and comment on future directions.

\section{Partition function in Fermi Gas formalism}
In this section we provide the Fermi gas formalism for the general $R$-charge assignments \eqref{Rgen}.
In the derivation we use the difference expression of the Chern-Simons levels \eqref{kins}, but not the explicit values of $s_a$.
The Fermi gas formalism hold not only for $s_a=\pm 1$ but also for arbitrary choices of their values, which correspond to the general ${\cal N}=3$ circular quiver superconformal Chern-Simons theories.

With the help of the localization technique, the partition function of this theory reduces into the following matrix model \cite{KWY1,KWY2,J,HHL}
\begin{align}
Z(N)=\frac{1}{(N!)^M}\prod_{a=1}^M\prod_{i=1}^N\int D\lambda_{a,i}\prod_{a=1}^M
\frac{
\prod_{i>j}2\sinh\frac{\lambda_{a,i}-\lambda_{a,j}}{2}
\prod_{i>j}2\sinh\frac{\lambda_{a+1,i}-\lambda_{a+1,j}}{2}
}
{\prod_{i,j}2\cosh\frac{\lambda_{a,i}-\lambda_{a+1,j}-\pi i\zeta_a}{2}},
\end{align}
where
\begin{align}
D\lambda_{a,i}=\frac{d\lambda_{a,i}}{2\pi}\exp\biggl[\frac{ik_a}{4\pi}\lambda_{a,i}^2\biggr]
\end{align}
with $k_a$ the Chern-Simons level on the $a$-th vertex given by \eqref{kins}.\footnote{
If we take $\zeta_a$ to be pure imaginary, this matrix model completely coincide to that with the mass deformations, whose large $N$ limit was studied for real Chern-Simons level $k$ in \cite{NST} and for complex $k$ in \cite{AR,AZ}.
}
Compared with the superconformal case $\zeta_a=0$, the only difference is the shift in the arguments of the cosine-hyperbolic factors which come from the 1-loop determinant of the bifundamental hypermultiplets.
This fact allows the straightforward application of the computational techniques in \cite{MP1} to derive the Fermi gas formalism.\footnote{
The Fermi gas formalism for a mass-deformed $U(N)\times U(N)$ theory with fundamental matter multiplets was also constructed in \cite{DF}.
}
First we rewrite the partition function as
\begin{align}
Z(N)=\frac{1}{N!}\prod_{i=1}^N\int\frac{d\lambda_{1,i}}{2\pi}\det_{i,j}\rho_0(\lambda_{1,i},\lambda_{1,j}),
\label{Zdet}
\end{align}
where
\begin{align}
\rho_0(v,w)&=\prod_{a=2}^M\int\frac{dz_a}{2\pi}
\biggl[e^{\frac{iks_1v^2}{8\pi}}\frac{1}{2\cosh\frac{v-z_2-i\pi\zeta_1}{2}}e^{-\frac{iks_1z_2^2}{8\pi}}\biggr]
\biggl[e^{\frac{iks_2z_2^2}{8\pi}}\frac{1}{2\cosh\frac{z_2-z_3-i\pi\zeta_2}{2}}e^{-\frac{iks_2z_3^2}{8\pi}}\biggr]\nonumber \\
&\quad\quad\quad\quad\quad \cdots
\biggl[e^{\frac{iks_Mz_M^2}{8\pi}}\frac{1}{2\cosh\frac{z_M-w-i\pi\zeta_2}{2}}e^{-\frac{iks_Mw^2}{8\pi}}\biggr].
\label{rho0}
\end{align}
The expression \eqref{Zdet} can be derived with the help of the Cauchy determinant formula
\begin{align}
\frac{\prod_{i<j}2\sinh\frac{x_i-x_j}{2}\prod_{i<j}2\sinh\frac{y_i-y_j}{2}}{\prod_{i,j}2\cosh\frac{x_i-y_j-\Delta}{2}}=\det_{i,j}\frac{1}{2\cosh\frac{x_i-y_j-\Delta}{2}}
\end{align}
and the formula (see appendix A in \cite{MM})
\begin{align}
\frac{1}{N!}\int dz^N\Bigl[\det_{i,j}f(x_i,z_j)\Bigr]\Bigl[\det_{i,j}g(z_i,y_j)\Bigr]=\det_{i,j}\Bigl[\int dzf(x_i,z)g(z,y_j)\Bigr].
\end{align}
Using the Fourier transformation formula
\begin{align}
\frac{1}{2\cosh\frac{z-\pi i\zeta}{2}}=\int \frac{dp}{2\pi}e^{\frac{ipz}{2\pi}}\frac{e^{\frac{\zeta p}{2}}}{2\cosh\frac{p}{2}},
\end{align}
each factor in the square bracket can be rewritten as
\begin{align}
e^{\frac{iks_az_a^2}{8\pi}}\frac{1}{2\cosh\frac{z_a-z_{a+1}-i\pi\zeta_2}{2}}e^{-\frac{iks_az_{a+1}^2}{8\pi}}
=k\cdot \biggl\langle x=kz_a\biggl|e^{\frac{is_a{\widehat x}^2}{8\pi k}}\frac{e^{\frac{\zeta{\widehat p}}{2}}}{2\cosh\frac{\widehat p}{2}}e^{-\frac{is_a{\widehat x}^2}{8\pi k}}\biggr|x=kz_{a+1}\biggr\rangle,
\label{op}
\end{align}
where we have introduced the canonical position/momentum operators $({\widehat x},{\widehat p})$ and their eigenstates $(|x\rangle,|p\rangle)$ normalized so that
\begin{align}
[{\widehat x},{\widehat p}]&=i\hbar,\quad(\hbar=2\pi k)\nonumber \\
\langle x|x^\prime\rangle=2\pi\delta(x-x^\prime),\quad
\langle p|p^\prime\rangle&=2\pi\delta(p-p^\prime),\quad
\langle x|p\rangle=\frac{1}{\sqrt{k}}e^{\frac{ipx}{2\pi k}}.
\end{align}
In the operator formalism, the $(M-1)$ integrations in $\rho_0$ \eqref{rho0} together with the $k$ factored out in \eqref{op} are interpreted as the insertion of unity
\begin{align}
1=\int \frac{dx}{2\pi}|x\rangle\langle x|,\quad (x=kz)
\end{align}
hence the partition function \eqref{Zdet} can be written as
\begin{align}
Z(N)=\frac{1}{N!}\prod_{i=1}^N\int\frac{dx_i}{2\pi}\det_{i,j}\langle x_i|{\widehat\rho}|x_j\rangle 
\end{align}
with
\begin{align}
{\widehat \rho}=
\frac{e^{\frac{\zeta_1}{2}({\widehat p}-\frac{s_1}{2}{\widehat x})}}{2\cosh[\frac{1}{2}({\widehat p}-\frac{s_1}{2}{\widehat x})]}
\frac{e^{\frac{\zeta_2}{2}({\widehat p}-\frac{s_2}{2}{\widehat x})}}{2\cosh[\frac{1}{2}({\widehat p}-\frac{s_2}{2}{\widehat x})]}
\cdots
\frac{e^{\frac{\zeta_M}{2}({\widehat p}-\frac{s_M}{2}{\widehat x})}}{2\cosh[\frac{1}{2}({\widehat p}-\frac{s_M}{2}{\widehat x})]},
\end{align}
where we have used \eqref{op} and the formula
\begin{align}
e^{\frac{i}{2\hbar}{\widehat x}^2}f({\widehat p})e^{-\frac{i}{2\hbar}{\widehat x}^2}=f({\widehat p}-{\widehat x}).
\end{align}
Using the Fredholm determinant formula, the grand potential \eqref{ZtoJ} can be written as
\begin{align}
J(\mu)=\Tr\log(1+e^\mu{\widehat \rho}).
\label{trlog}
\end{align}
This is the same form as the grand potential of a quantum statistical system of the ideal Fermi gas.

As in the superconformal case, a special simplification occurs if the original theory have the ${\cal N}=4$ supersymmetry \eqref{sas}.
Since $s_a$ takes $\pm 1$ in this case there are only two kinds of argument in the density matrix
\begin{align}
{\widehat Q}=-{\widehat p}+\frac{{\widehat x}}{2},\quad
{\widehat P}={\widehat p}+\frac{{\widehat x}}{2},\quad
([{\widehat Q},{\widehat P}]=i\hbar).
\end{align}

In the remaining part of this paper, we further focus on the class of the minimal separation of $s_a=\pm 1$ \eqref{sminimal} where the (hermitized) density matrix is
\begin{align}
{\widehat \rho}=
\frac{e^{\frac{\xi{\widehat Q}}{4}}}{\bigl(2\cosh\frac{\widehat Q}{2}\bigr)^{\frac{q}{2}}}
\frac{e^{\frac{\eta{\widehat P}}{2}}}{\bigl(2\cosh\frac{\widehat P}{2}\bigr)^p}
\frac{e^{\frac{\xi{\widehat Q}}{4}}}{\bigl(2\cosh\frac{\widehat Q}{2}\bigr)^{\frac{q}{2}}}.
\end{align}
with $\xi$ and $\eta$ given as \eqref{xieta}.
Since $\zeta_a$ on each edge is bounded as \eqref{zetabound}, $\xi$ and $\eta$ are bounded as
\begin{align}
-q<\xi<q,\quad -p<\eta<p.
\label{xietabound}
\end{align}
This ensures that the density matrix decays at the infinity of the phase space and thus the trace $\Tr$ in \eqref{trlog} is well defined.

\section{Perturbative expansion in $1/N$}
\label{Fermisurface}
In this section we show that the large $\mu$ expansion of the grand potential $J(\mu)$ takes the form of \eqref{Jpert}, with $C,B$ and $A$ given as \eqref{CBA}, up to the non-perturbative corrections ${\cal O}(e^{-\mu})$.
Here $C,B$ and $A$ are $\mu$-independent constants given as \eqref{CBA}.
Plugging these expressions into the inversion formula \eqref{ZtoJ}, we obtain the all order perturbative expansion of the partition function in $1/N$, which sum up to an Airy function as \eqref{Airy}.

As argued in \cite{MP1}, the perturbative expansion of $J(\mu)$ \eqref{Jpert} follows from the large $E$ expansion of the number of states $n(E)$ with energy below $E$
\begin{align}
n(E)=\Tr\theta(E-{\widehat H})=CE^2+B-\frac{\pi^2C}{3}+{\cal O}(e^{-E}),
\label{n}
\end{align}
where ${\widehat H}$ is the Hamiltonian operator given by the logarithm of the density matrix:
\begin{align}
e^{-{\widehat H}}=e^{-U({\widehat Q})/2}e^{-T({\widehat P})}e^{-U({\widehat Q})/2}
\label{Hhat}
\end{align}
with
\begin{align}
U({\widehat Q})=q\log\biggl[2\cosh\frac{\widehat Q}{2}\biggr]-\frac{\xi{\widehat Q}}{2},\quad 
T({\widehat P})=p\log\biggl[2\cosh\frac{\widehat P}{2}\biggr]-\frac{\eta{\widehat P}}{2}.
\label{UT}
\end{align}
Below we shall derive the behavior \eqref{n} as well as the explicit expressions for $C$ and $B$.
On the other hand, the overall constant $A$ requires a non-perturbative analysis of the grand potential and treated in the next section.

First of all, we introduce the Wigner transformation $({\widehat X})_W$ of an arbitrary operator ${\widehat X}$
\begin{align}
({\widehat X})_W=\int\frac{dQ^\prime}{2\pi}\biggl\langle Q-\frac{Q^\prime}{2}\biggl|{\widehat X}\biggr|Q+\frac{Q^\prime}{2}\biggr\rangle e^{\frac{iQ^\prime P}{\hbar}}.
\end{align}
Then $n(E)$ is approximately given by the volume inside the region $F=\{(Q,P)\in\mathbb{R}^2|H_W\le E\}$ divided by $2\pi\hbar$ as\footnote{
The approximation ``$\approx$'' in \eqref{nE} is due to the fact that $f({\widehat {\cal O}})_W\neq f({\cal O}_W)$ in general.
The deviation, however, is irrelevant to the perturbative expansion \eqref{n} as argued in \cite{MP2,MN1}.
}
\begin{align}
n(E)\approx \int\frac{dQdP}{2\pi\hbar}\theta(E-H_W).
\label{nE}
\end{align}
In the limit of $E\rightarrow\infty$ we can approximate the Wigner Hamiltonian $H_W$ with the classical Hamiltonian
\begin{align}
H_0=U(Q)+T(P)
\label{Hcl}
\end{align}
and further approximate the functions $U(Q)$ and $T(P)$ as $U(Q)\approx (q|Q|-\xi Q)/2$, $T(P)\approx (p|P|-\eta P)/2$.
In this limit the region $F$ approaches a polygon
\begin{align}
F_\text{pol}=\biggl\{(Q,P)\in\mathbb{R}^2\biggl|\frac{q|Q|-\xi Q}{2}+\frac{p|P|-\eta P}{2}\le E\biggr\}
\end{align}
and the leading part of $n(E)$ is straightforwardly obtained as
\begin{align}
n(E)=CE^2+\delta n
\end{align}
with $C$ given by \eqref{CBA}.

\begin{figure}[!t]
\begin{center}
\includegraphics[width=9cm]{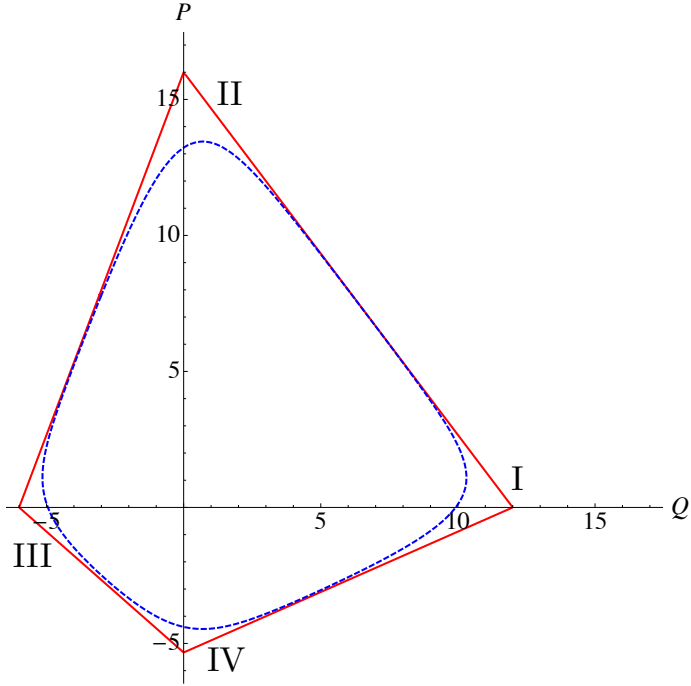}
\caption{
The boundary of the region $F$ with $(q,\xi;p,\eta,E)=(1,1/3;1,1/2,4)$ (dashed blue line) and that of the polygon $F_\text{pol}$ (solid red line).
%
%
\label{figvol}
}
\end{center}
\end{figure}

To compute the correction $\delta n$, we have to take into account the two effects which deform the boundary of $F$ from that of the polygon $F_\text{pol}$: (i) the deviation of the Wigner Hamiltonian from the classical Hamiltonian $H_0$ \eqref{Hcl}, and (ii) the deviation of $U(Q)$ and $T(P)$ \eqref{UT} from the linear functions.
First consider the deviation (i).
The Wigner Hamiltonian can be computed order by order in $\hbar$, by first computing the Hamiltonian operator \eqref{Hhat} and then performing the Wigner transformation using the formulas
\begin{align}
f({\widehat Q})_W=f(Q),\quad f({\widehat P})_W=f(P),\quad ({\widehat X}{\widehat Y})_W=X_W\star Y_W,
\end{align}
where $\star$ is the non-commutative product
\begin{align}
\star=\exp\biggl[\frac{i\hbar}{2}\Bigl(\overleftarrow{\partial}\!_Q\overrightarrow{\partial}\!_P-\overleftarrow{\partial}\!_P\overrightarrow{\partial}\!_Q\Bigr)\biggr].
\end{align}
Notice that the second derivatives of $U(Q)$ and $T(P)$ are exponentially suppressed for large arguments.
Therefore, since at least one of $Q$ and $P$ is of order $E$ on the boundary of the polygon $F_\text{pol}$, we can neglect all the terms containing $(\partial^m_QU)(\partial^n_PT)$ with $m,n\ge 2$ for the purpose to compute the deviation $\delta n$ perturbatively in $1/E$, and the Wigner Hamiltonian can be approximated with
\begin{align}
H_W=U+T+\frac{\hbar^2}{24}(U^\prime)^2T^{(2)}-\frac{\hbar^2}{12}U^{(2)}(T^\prime)^2+\sum_{\ell\ge 3}(c^{(\ell)}_U(U^\prime)^\ell T^{(\ell)}+c^{(\ell)}_T(T^\prime)^\ell U^{(\ell)})+\cdots,
\end{align}
where $c_U^{(\ell)}$ and $c_T^{(\ell)}$ are some constants, while $U^{(\ell)}=\partial^\ell_QU$ and $T^{(\ell)}=\partial^\ell_PT$.
The boundary $H_W(Q,P)=E$ of the region $F$ is displayed in Figure \ref{figvol}.
The deformation of the surface is negligible except around the four corners of the polygon where the deviation (ii) is relevant.
To compute $\delta n$ we shall decompose it into the contributions around each corner
\begin{align}
\delta n=-\frac{1}{2\pi\hbar}(vol(\text{I})
+vol(\text{II})
+vol(\text{III})
+vol(\text{IV})
).
\label{deltavol}
\end{align}
First let us consider the region I.
Since $Q\sim E$ in this region, we can replace $U(Q)\rightarrow (q-\xi)Q/2$ in our calculation without loss of any perturbative corrections.
Under this approximation the Fermi surface adjacent to the region I is characterized as
\begin{align}
E=\frac{q-\xi}{2}Q+T+\frac{\hbar^2(q-\xi)^2}{96}T^{(2)}+\sum_{\ell\ge 3}c^{(\ell)}_U\Bigl(\frac{q-\xi}{2}\Bigr)^\ell T^{(\ell)}.
\end{align}
Denoting the points on the boundary of $F$ as $(Q_F(P),P)$ while those on the boundary of the polygon $(Q_{\text{pol}}(P),P)$, we can compute the volume of the region I as
\begin{align}
vol(\text{I})&=\int_{P_-}^{P_+}dP(Q_{\text{pol}}-Q_F)\nonumber \\
&=\frac{2}{q-\xi}\int_{P_-}^{P_+}\biggl(T-\frac{p|P|-\eta P}{2}+\frac{\hbar^2(q-\xi)^2}{96}T^{(2)}+\sum_{\ell\ge 3}c^{(\ell)}_U\Bigl(\frac{q-\xi}{2}\Bigr)^\ell T^{(\ell)}\biggr).
\label{volI}
\end{align}
Here $P_\pm$ correspond to some upper/lower bound: the midpoints of the edges of the approaching polygon for instance.
Since $P_\pm\sim E$, at the perturbative level we can replace them with $\pm\infty$ as the integrand in \eqref{volI} is exponentially suppressed for large $P$, to obtain
\begin{align}
vol(\text{I})=\frac{\pi^2p}{3(q-\xi)}+\frac{\pi^2k^2p(q-\xi)}{12}.
\end{align}
Note that the last terms in \eqref{volI} do not contribute as they give vanishing boundary terms.
Similarly we can evaluate the volume of region II, III and IV as
\begin{align}
vol(\text{II})&=\frac{\pi^2q}{3(p-\eta)}-\frac{\pi^2k^2q(p-\eta)}{6},\quad
vol(\text{III})=\frac{\pi^2p}{3(q+\xi)}+\frac{\pi^2k^2p(q+\xi)}{12},\nonumber \\
vol(\text{IV})&=\frac{\pi^2q}{3(p+\eta)}-\frac{\pi^2k^2q(p+\eta)}{6}.
\end{align}
Substituting these results into $\delta n$ \eqref{deltavol}, we finally obtain the expression of $B$ in \eqref{CBA}.

\section{Non-perturbative effects in grand potential}
\label{WKB}
In this section we study the non-perturbative corrections to the grand potential which we shall call the instantons, as well as the constant $A$ in the perturbative part \eqref{Jpert}.

To evaluate the large $\mu$ expansion of the grand potential systematically, we shall use the following Mellin-Barnes expression of the grand potential $(\epsilon>0)$ \cite{H}
\begin{align}
J(\mu)=-\int^{\epsilon+i\infty}_{\epsilon-i\infty}\frac{dt}{2\pi i}\Gamma(t)\Gamma(-t){\cal Z}(t)e^{t\mu}
\label{MB}
\end{align}
with
\begin{align}
{\cal Z}(t)=\Tr e^{-t{\widehat H}}.
\end{align}
If we assume $\mu<0$ the right-hand side of \eqref{MB} can be evaluated by pinching the integration contour so that it surrounds the right-half of the whole complex plane $\mathbb{C}$.
Collecting the residues in this region, i.e. the residues at $t=1,2,\cdots $ we indeed obtain the small $e^\mu$ expansion of the original expression \eqref{trlog}.
On the other hand, if $\mu>0$ we can pinch the contour so that it surrounds the left-half of $\mathbb{C}$.
As a result, the grand potential is expressed as the sum of the residues over $\mathrm{Re}(t)\le 0$.
Due to the factor $e^{t\mu}$ in the integrand the residues are typically small or at most polynomial for large $\mu$, which immediately give the large $\mu$ expansion of the grand potential.

We can compute ${\cal Z}(n)$ order by order in the small $\hbar$ expansion
\begin{align}
{\cal Z}(n)=\sum_{s=0}^\infty\hbar^{2s-1}{\cal Z}_{2s}(n)
\label{calZWKB}
\end{align}
by the similar calculation as in the case of $\xi=\eta=0$ performed in \cite{MN1,MN2}.
Indeed, the only difference between the density matrix ${\widehat \rho}=e^{-{\widehat H}}$ with ${\widehat H}$ \eqref{Hhat} and that for $\xi=\eta=0$ is the definition of the unmixed operators $U({\widehat Q})$ and $T({\widehat P})$.
After the tedious calculation, we have obtained ${\cal Z}_0$, ${\cal Z}_2$, ${\cal Z}_4$ and ${\cal Z}_6$, which are displayed in appendix \ref{apcalZ}.

\subsection{$A$ in the perturbative part}
Before going on to the non-perturbative part, let us look the perturbative part of the grand potential again.
In the Mellin-Barnes representation \eqref{MB}, this comes from the residue at $t=0$.
From the explicit expression of ${\cal Z}_{2s}(n)$ \eqref{calZs} we obtain
\begin{align}
J^\text{pert}(\mu)=\frac{C}{3}\mu^3+B\mu+A.
\end{align}
Here $C$ and $B$ are constants which we have already computed in section \ref{Airy}, and $A$ is
\begin{align}
A&=
\frac{qp(q^3-q\xi^2+p^3-p\eta^2)\zeta(3)}{\pi^2(q^2-\xi^2)(p^2-\eta^2)k}
-\frac{qp(q+p)k}{24}
-\frac{\pi^2qp(qp(q+p)+3p\xi^2+3q\eta^2)k^3}{8640}\nonumber \\
&+\frac{\pi^4qp(qp(q^3+p^3)+5(p\xi^4+q\eta^4)+10qp(q\xi^2+p\eta^2))k^5}{1814400}
+{\cal O}(k^7).
\end{align}
At first sight the expression looks complicated.
With the simple decomposition structure we conjectured in the superconformal case \cite{MN1} in mind, however, we figure out the following decomposition structure again in this case
\begin{align}
A=\frac{p^2}{4}(f((q+\xi)k)+f((q-\xi)k))+\frac{q^2}{4}(f((p+\eta)k)+f((p-\eta)k)),
\label{Adecompose}
\end{align}
where $f(k)$ is given in a series expansion as
\begin{align}
f(k)=\frac{2\zeta(3)}{\pi^2 k}-\frac{k}{12}-\frac{\pi^2k^3}{4320}+\frac{\pi^4k^5}{907200}+{\cal O}(k^7).
\end{align}
The series $f(k)$ coincide with the small $k$ expansion of the constant map in the ABJM theory $A_\text{ABJM}(k)$.
Indeed once the structure \eqref{Adecompose} is postulated, we can deduce that $f(k)=A_\text{ABJM}(k)$ by taking the limit $(q,\xi;p,\eta)\rightarrow (1,0;1,0)$ where our theory reduces to the ABJM theory.
From these observations we conjecture the exact expression of $A$ for finite $k$ as \eqref{CBA}.
The conjecture is also confirmed from the exact computations of the partition function for $k\in\mathbb{N}$ in section \ref{finitek}.

\subsection{Instantons}
Due to the factor $e^{t\mu}$ in the Mellin-Barnes representation \eqref{MB}, all the residues at the poles with $\mathrm{Re}(t)<0$ are exponentially suppressed in $\mu$.
In this section we consider these non-perturbative effects in the grand potential, which we shall call the membrane instantons in an analogy of the ABJM case.

First we observe the following universal structure of ${\cal Z}_{s}(n)$
\begin{align}
{\cal Z}_{s}(n)=f_{s}(n)\times {\cal Z}_0(n),
\end{align}
where $f_{s}(n)$ are some rational functions of $n$.
Each $f_s(n)$ have at most a finite number of poles, all of which are cancelled with the zeroes of ${\cal Z}_0(n)$ at the same $n$.
From this structure it follows that the instanton species are independent of the order of the small $\hbar$ expansion.
Here we shall display only the ${\cal O}(\hbar^{-1})$ part of the non-perturbative part of the grand potential,
\begin{align}
J^\text{np}(\mu)=\frac{1}{\hbar}J_0^\text{np}+{\cal O}(\hbar)
\end{align}
with
\begin{align}
J_0^{np}=
\sum_{n=1}^\infty c^{(1)}_ne^{-\frac{2n\mu}{q+\xi}}
+\sum_{n=1}^\infty c^{(2)}_ne^{-\frac{2n\mu}{q-\xi}}
+\sum_{n=1}^\infty c^{(3)}_ne^{-\frac{2n\mu}{q+\eta}}
+\sum_{n=1}^\infty c^{(4)}_ne^{-\frac{2n\mu}{q-\eta}}
+\sum_{n=1}^\infty c^{(5)}_ne^{-n\mu}.
\label{mbinst}
\end{align}
The instanton coefficients are
\begin{align}
c^{(1)}_n&=\frac{(-1)^{n-1}}{\pi n!(q+\xi)}
\frac{
\Gamma\bigl(-\frac{2n}{q+\xi}\bigr)
\Gamma\bigl(\frac{2n}{q+\xi}\bigr)
\Gamma\bigl(-\frac{q-\xi}{q+\xi}n\bigr)
\Gamma\bigl(-\frac{p+\eta}{q+\xi}n\bigr)
\Gamma\bigl(-\frac{p-\eta}{q+\xi}n\bigr)
}
{
\Gamma\bigl(-\frac{2q}{q+\xi}n\bigr)
\Gamma\bigl(-\frac{2p}{q+\xi}n\bigr)
},\nonumber \\
c^{(2)}_n&=\frac{(-1)^{n-1}}{\pi n!(q-\xi)}
\frac{
\Gamma\bigl(-\frac{2n}{q-\xi}\bigr)
\Gamma\bigl(\frac{2n}{q-\xi}\bigr)
\Gamma\bigl(-\frac{q+\xi}{q-\xi}n\bigr)
\Gamma\bigl(-\frac{p+\eta}{q-\xi}n\bigr)
\Gamma\bigl(-\frac{p-\eta}{q-\xi}n\bigr)
}
{
\Gamma\bigl(-\frac{2q}{q-\xi}n\bigr)
\Gamma\bigl(-\frac{2p}{q-\xi}n\bigr)
},\nonumber \\
c^{(3)}_n&=\frac{(-1)^{n-1}}{\pi n!(p+\eta)}
\frac{
\Gamma\bigl(-\frac{2n}{p+\eta}\bigr)
\Gamma\bigl(\frac{2n}{p+\eta}\bigr)
\Gamma\bigl(-\frac{p-\eta}{p+\eta}n\bigr)
\Gamma\bigl(-\frac{q+\xi}{p+\eta}n\bigr)
\Gamma\bigl(-\frac{q-\xi}{q+\eta}n\bigr)
}
{
\Gamma\bigl(-\frac{2p}{p+\eta}n\bigr)
\Gamma\bigl(-\frac{2q}{p+\eta}n\bigr)
},\nonumber \\
c^{(4)}_n&=\frac{(-1)^{n-1}}{\pi n!(p-\eta)}
\frac{
\Gamma\bigl(-\frac{2n}{p-\eta}\bigr)
\Gamma\bigl(\frac{2n}{p-\eta}\bigr)
\Gamma\bigl(-\frac{p+\eta}{p-\eta}n\bigr)
\Gamma\bigl(-\frac{q+\xi}{p-\eta}n\bigr)
\Gamma\bigl(-\frac{q-\xi}{q-\eta}n\bigr)
}
{
\Gamma\bigl(-\frac{2p}{p-\eta}n\bigr)
\Gamma\bigl(-\frac{2q}{p-\eta}n\bigr)
},\nonumber \\
c^{(5)}_n&=-\frac{(-1)^{n-1}}{2\pi n}
\frac{
\Gamma\bigl(-\frac{q+\xi}{2}n\bigr)
\Gamma\bigl(-\frac{q-\xi}{2}n\bigr)
\Gamma\bigl(-\frac{p+\eta}{2}n\bigr)
\Gamma\bigl(-\frac{p-\eta}{2}n\bigr)
}
{
\Gamma(-qn)
\Gamma(-pn)
},
\label{instcoefs}
\end{align}
where we have used the expression for the Euler beta function $\B(x,y)$ in the Gamma functions \eqref{Beta} to clarify the pole structure of the instanton coefficients.

\subsection{Divergence and mixing of instantons}
Lastly let us study the divergent structure of the instantons.
Respecting the original setup of the quiver Chern-Simons theory here we assume $q,p\in\mathbb{N}$.

First we consider the fifth kind of the instanton.
Since the coefficients of this instanton $c^{(5)}_n$ contain the divergent factors $\Gamma(-qn)\Gamma(-pn)$ in the denominator, the coefficients generically vanishes for all $n\ge 1$.
The only exception happens if the arguments of the Gamma functions in the denominator are also negative integers so that the divergences in the numerator compensate the divergences in the denominator.
In those case the exponent coincides with that of some instanton in the other four kinds and the HMO pole cancellation mechanism \cite{HMO2,MN2} occurs.
In this sense, these fifth kind of instantons never produce distinctive instanton effects, as called the ``ghost instantons'' in \cite{MN2}.

Next we consider the other four kinds of instantons.
In the superconformal limit $\xi=\eta=0$ they reduce to the two kinds of the membrane instantons $(e^{-2\mu/q}, e^{-2\mu/p})$ which would be associated with the $\mathbb{Z}_q$- and $\mathbb{Z}_p$-orbifold in the background geometry.
Indeed each instanton coefficient \eqref{instcoefs} have similar structure individually.
The rules for the divergence and the mixing are, however, slightly complicated than those in the superconformal case:
\begin{itemize}
\item{In the superconformal case, the coefficient always diverges when the instanton exponent coincide with that of another instanton.
This divergence is cancelled by the divergence of the other instanton with the same exponent.
For the mixing among the four instantons in the current case, this is not alway the case.
When the mixing is between only the first two ($e^{-2\mu/(q+\xi)},e^{-2\mu/(q-\xi)})$ or the last two ($e^{-2\mu/(p+\eta)},e^{-2\mu/(p-\eta)})$ the individual coefficients remain finite.
}
\item{In the superconformal case, the mixing and the pole cancellation are inevitable, since $q,p\in\mathbb{N}$ as obvious from their roles in the orbifold.
Due to this restriction the slight deformation such as $q\rightarrow q+\epsilon$ to disentangle the mixing pair is unphysical.
In the current case, however, $\xi,\eta$ are continuous parameters of the original theory.
This suggests that, not only the finite part remaining after the cancellation but also the divergence itself would have some gravitational counterpart.
}
\end{itemize}

\section{Exact partition function for finite $(k,N)$}
\label{finitek}
In the superconformal case $\xi=\eta=0$, a particular structure of the density matrix allows the systematic computation of the partition function with finite $k,N\in\mathbb{N}$.
As we will see below, the method can be generalized for the case without superconformal symmetry if $\xi$ and $\eta$ are rational numbers.
In this section we concentrate on the deformation of the ABJM theory, $q=p=1$ and compute the partition function for various $k,N\in\mathbb{N}$ and $\xi,\eta\in\mathbb{Q}$.

\subsection{Systematic computation of partition function}
In this section we display the algorithm to compute the traces of the density matrix $\Tr{\widehat \rho}^n$ for given $k,\xi,\eta$ recursively in $n$.
The partition functions $Z(N)$ can be read off through the definition of the grand potential \eqref{ZtoJ} and \eqref{trlog}.
Here we would like to consider only the case with $q=p=1$.
The way to extend the method for the case with general $q,p\in\mathbb{N}$ is identical to that in the superconformal case \cite{MN3,HHO}.

The essence for this method is the following schematic structure of the density matrix
\begin{align}
\rho(Q_1,Q_2)=\frac{1}{2\pi}\langle Q_1|{\widehat \rho}|Q_2\rangle =\frac{E(Q_1)E(Q_2)}{\alpha A(Q_1)+\alpha^{-1}A(Q_2)}.
\label{rhosch}
\end{align}
The explicit expression of each ingredient is
\begin{align}
E(Q)=\frac{e^{(\frac{\xi}{4}+\frac{1}{2k})Q}}{\bigl(2\cosh\frac{Q}{2}\bigr)^{\frac{1}{2}}},\quad
A(Q)=2\pi ke^{\frac{Q}{k}},\quad \alpha=e^{-\frac{\pi i\eta}{2}}.
\end{align}
From the structure \eqref{rhosch} the powers of the density matrix are
\begin{align}
\rho^n(Q_1,Q_2)=\frac{1}{2\pi}\langle Q_1|{\widehat \rho}^n|Q_2\rangle =\frac{E(Q_1)E(Q_2)}{A(Q_1)-(-1)^n\alpha^{-2n}A(Q_2)}\sum_{m=0}^{n-1}(-1)^m\alpha^{-2m-1}\psi_m(Q_1)\phi_{n-m-1}(Q_2)
\end{align}
with
\begin{align}
\psi_m(Q)=\frac{1}{E(Q)}\int dQ^\prime \rho^m(Q,Q^\prime)E(Q^\prime),\quad
\phi_m(Q)=\int dQ^\prime E(Q^\prime)\rho^m(Q^\prime,Q)\frac{1}{E(Q)}.
\end{align}
Note that $\phi_m(Q)$ are related to $\psi_m(Q)$ through the complex conjugation or the replacement $\alpha\rightarrow \alpha^{-1}$.
Now the task to compute the powers of a matrix $\rho^n(Q_1,Q_2)$ is reduced into the computation of the vectors $\psi_n(Q)$ and $\phi_n(Q)$.
Moreover, these vectors can be computed by simple iterative steps if $\xi\in\mathbb{Q}$, as we shall see below.

We would like to focus on the vector $\psi_m(Q)$ which obeys the recursion relation
\begin{align}
\psi_{m+1}(Q)=\frac{1}{E(Q)}\int dQ^\prime \rho(Q,Q^\prime)E(Q^\prime)\psi_m(Q^\prime).
\label{recur}
\end{align}
For $\xi$ being a rational number, say $\xi=b/a$ for some $a,b\in\mathbb{N}$ ($a\perp b$), we can introduce a new integration variable $u=e^{Q/w}$ with $w=\lcm(2a,k)$ to rewrite the recursion relation \eqref{recur} as
\begin{align}
\psi_{m+1}(u)=\frac{x\alpha}{2\pi}\int_0^\infty dv\frac{1}{v^x+\alpha^2u^x}\frac{v^{x+y+\frac{w}{2}-1}}{v^w+1}\psi_m(v).
\end{align}
Here $x$ and $y$ are integers given by $x=w/k$ and $y=w\xi/2$.
This integration relation is in the same type as that used in the ABJM theory \cite{PY} and can be simplified by expanding $\psi_m(u)$ in the series of $\log u$
\begin{align}
\psi_m(u)=\sum_{j\ge 0}\psi^{(j)}_m(u)(\log u)^j
\end{align}
with $\psi^{(j)}_m(u)$ rational functions in $u$, as
\begin{align}
\psi_m(u)=
-\frac{x\alpha}{2\pi}
\sum_{j\ge 0}\frac{(2\pi i)^{j+1}}{j+1}
\sum_{v_a\in\mathbb{C}\backslash \mathbb{R}^+}
\Res\biggl[
\frac{1}{v^x+\alpha^2u^x}\frac{v^{x+y+\frac{w}{2}-1}}{v^w+1}\psi^{(j)}_m(v)B_{j+1}\biggl[\frac{\log^{(+)} v}{2\pi i}\biggr],v\rightarrow v_a\biggr].
\label{psiit}
\end{align}
Here $\log^{(+)}$ is the logarithm function with the branch cut on $\mathbb{R}^+$, and $B_j(z)$ are the Bernoulli polynomials.
In the contribution of the poles associated with the first factor $1/(v^x+\alpha^2u^x)$, we assume $u\in\mathbb{R}^+$.

Once $\psi_m(u)$ and $\phi_m(u)$ are computed in this manner, the integration in the computation of $\Tr{\widehat \rho}^n$ from \eqref{rhosch} can be manipulated in the same way, and we finally obtain, when $(-1)^n\alpha^{-2n}\neq 1$
\begin{align}
\mathrm{Tr}{\widehat \rho}^n=-\frac{x}{2\pi}
\frac{1}{1-(-1)^n\alpha^{-2n}}\sum_{j\ge 0}\frac{(2\pi i)^{j+1}}{j+1}\Res\biggl[\sum_{v_a\in\mathbb{C}\backslash \mathbb{R}^+}\frac{v^{y+\frac{w}{2}-1}}{v^w+1}f^{(j)}_n(v)B_{j+1}\biggl[\frac{\log^\prime v}{2\pi i}\biggr],v\rightarrow v_a\biggr],
\label{trrhoTWPY1}
\end{align}
while for $(-1)^n\alpha^{-2n}=1$
\begin{align}
\mathrm{Tr}{\widehat \rho}^n=-\frac{1}{2\pi}
\sum_{j\ge 0}\frac{(2\pi i)^{j+1}}{j+1}\Res\biggl[\sum_{v_a\in\mathbb{C}\backslash \mathbb{R}^+}\frac{v^{y+\frac{w}{2}}}{v^w+1}g^{(j)}_n(v)B_{j+1}\biggl[\frac{\log^\prime v}{2\pi i}\biggr],v\rightarrow v_a\biggr].
\label{trrhoTWPY2}
\end{align}
Here $f^{(j)}_n(u)$ and $g^{(j)}_n(u)$ are the rational functions defined by
\begin{align}
\sum_{m=0}^{n-1}(-1)^m\alpha^{-2m-1}\psi_m(u)\phi_{n-m-1}(u)&=\sum_{j\ge 0}f^{(j)}_n(u)(\log u)^j,\nonumber \\
\sum_{m=0}^{n-1}(-1)^m\alpha^{-2m-1}\partial_u\psi_m(u)\phi_{n-m-1}(u)&=\sum_{j\ge 0}g^{(j)}_n(u)(\log u)^j.
\end{align}

\subsubsection{Poles generated by iterations}
In the case $\eta=0$, the poles we need to take into account in the iteration \eqref{psiit} are always the following two series
\begin{align}
v&=e^{\frac{\pi i(2\ell-1)}{x}}\alpha^{\frac{2}{x}}u,\quad (\ell=1,2,\cdots x)\nonumber \\
v&=e^{\pi i(2\ell-1)/w},\quad (\ell=1,2,\cdots w)
\label{defaultpoles}
\end{align}
where the poles in the first line come from the first factor $1/(v^x+\alpha^2u^x)$ in the right-hand side of \eqref{psiit} and the second line from the second factor $1/(v^w+1)$.

The situation is different for general values of $\eta$, since the third factor $\psi_m^{(j)}(v)$ in the right-hand side of \eqref{psiit} may have distinct poles.
These poles are generated by both the residue at the $u$-dependent poles and that for the $u$-independent poles.
To clarify the pole contents of $\psi_m(u)$,\footnote{
Here we do not mind the order of each pole.
}
first let us study the poles generated in the step $\psi_1\rightarrow \psi_2$.
From the residues at the poles in the first line of \eqref{defaultpoles}, we find
\begin{align}
\psi_2(u)&\propto \frac{1}{(\alpha^{\frac{2}{x}}ue^{\frac{\pi i(2j-1)}{x}})^w+1}\nonumber \\
&\propto \frac{1}{u^w+e^{\pi ik}\alpha^{-2k}}
\end{align}
where we have used the fact $w/x=k\in\mathbb{N}$.
On the other hand the poles in the second line in \eqref{defaultpoles} generates
\begin{align}
\psi_2(u)&\propto \frac{1}{(e^{\frac{\pi i(2\ell-1)}{w}})^x+\alpha^2u^x}\nonumber \\
&\propto \frac{1}{u^w+e^{\pi ik}\alpha^{-2k}},
\end{align}
where in the second line we have reduced together the fractions with $\ell=1,2,\cdots w$.
From these results we conclude that $\psi_2(u)$ have the new poles associated with the factor $1/(u^w+e^{\pi ik}\alpha^{-2k})$.
In the step $\psi_2\rightarrow\psi_3$ the cross substitution of the poles of this factor and those of the first factor in \eqref{psiit} again generates the new pole factors $1/(u^w+\alpha^{-4k})$.
Repeating these arguments we conclude that $\psi_m(u)$ have the following poles
\begin{align}
\psi_m(u)\propto\prod_{\ell=0}^{m-1}\frac{1}{u^w+e^{\pi ik\ell}\alpha^{-2k\ell}}.
\end{align}
In the computation of $\psi_m(u)$ \eqref{psiit} and $\Tr{\widehat \rho}^n$ \eqref{trrhoTWPY1}, \eqref{trrhoTWPY2}, we need to take into account these poles.
\begin{table}[!t]
\begin{center}
\begin{tabular}{|r|rrr|}
\hline
$\xi\backslash\eta$&1/4&1/3&1/2\\
\hline
0                  &  4&  5&  8\\
1/4                &  3&  4&  5\\
1/3                & --&  2&  6\\
1/2                & --& --&  5\\
\hline
\end{tabular}
\end{center}
\caption{The values of $N_\text{max}$ for each $(k=4,\xi,\eta)$ in our computation.
We have chosen $\xi,\eta$ as $\xi\le \eta$ since the matrix model is trivially symmetric under $\xi\leftrightarrow\eta$.
}
\label{k4data}
\end{table}

\subsection{Comparison with small $k$ expansion}
In this section we compare the exact values with the results of the semiclassical analysis in section \ref{Fermisurface},\ref{WKB} to confirm the conjectural expression for $A$ \eqref{CBA} and observe the non-perturbative effects corresponding to the worldsheet instantons ${\cal O}(e^{-\frac{\mu}{k}})$.
For this purpose we have computed the exact partition functions for $k=4$ and various pairs of $(\xi,\eta)$, $Z_k^{(\xi;\eta)}(N)$ with $N=1,2,\cdots,N_\text{max}$, where the values of $N_\text{max}$ are listed in table \ref{k4data}.
The exact values are collected in appendix \ref{evlist}.

The expression of $A$ in \eqref{CBA} can be evaluated for finite $k$ by the integral expression of the constant map in the ABJM theory \cite{HaOk1}
\begin{align}
A_\text{ABJM}(k)=\frac{2\zeta(3)}{\pi^2k}\biggl(1-\frac{k^3}{16}\biggr)+\frac{k^2}{\pi^2}\int_0^\infty dx\frac{x}{e^{kx}-1}\log(1-e^{-2x}).
\label{Afinitek}
\end{align}
Comparing these values and those obtained by fitting the exact values of the partition function \eqref{Zexactsk41}-\eqref{Zexactsk49} with the Airy function \eqref{Airy} with $B$ and $C$ \eqref{CBA}, we confirm that our conjecture \eqref{CBA} is indeed correct (see table \ref{omegafit} for $k=4$).

\begin{figure}[!t]
\begin{center}
\includegraphics[width=9cm]{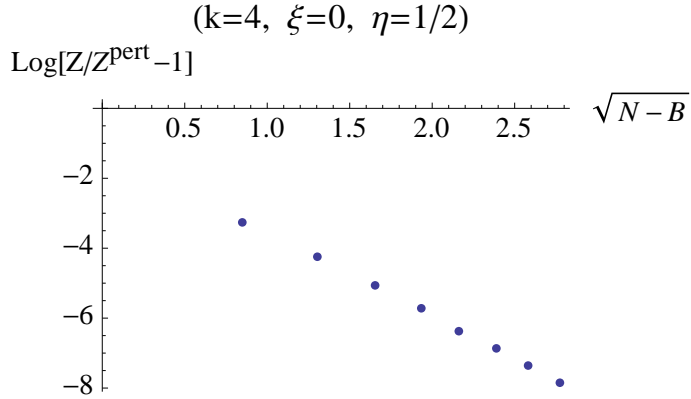}
\caption{
The non-perturbative part of the exact partition function remaining after the subtraction of the perturbative part $Z^\text{pert}(N)$ for $(k,\xi,\eta)=(4,0,1/2)$.
The combination $Z(N)/Z^\text{pert}(N)-1$ behaves like $e^{-\omega\sqrt{\frac{N-B}{C}}}$ even for small $N$.
}
\end{center}
\label{evinst}
\end{figure}

Next let us compare the instanton exponent.
With the help of the inversion formula \eqref{JtoZ}, the leading non-perturbative effect can be directly related to the exact values as \cite{PY}
\begin{align}
J(\mu)-J^\text{pert}(\mu)\sim e^{-\omega\mu}\Longleftrightarrow\frac{Z(N)}{Z^\text{pert}(N)}-1\sim e^{-\sqrt{\frac{N-B}{C}}}.
\end{align}
We observe that this approximation for the exact values holds even for small $N$ (see figure \ref{evinst}) and thus we can estimate the leading instanton exponents $\omega$ by the fitting as in table \ref{omegafit}.
Since the results are considerably different from the leading exponent $e^{-2\mu/(1+\eta)}$ obtained from the WKB expansion, we conclude that they correspond to the worldsheet instanton effects ${\cal O}(e^{-\frac{\mu}{k}})$.

\begin{table}[!t]
\begin{center}
\begin{tabular}{|rrr|rr|r||r|rr|}
\hline
$k$&$\xi$&$\eta$&$A$ (\eqref{CBA})&$A$ (fitting)&fitting/\eqref{CBA}&$\omega$(fitting)&$\frac{2}{1+\eta}$&$\frac{4}{k(1+\xi)(1+\eta)}$\\
\hline
$4$&$0$  &$1/4$&$-0.38131$&$-0.38134$&$1.00009$&$0.79$&$1.6$&$0.8$\\
$4$&$0$  &$1/3$&$-0.38442$&$-0.38443$&$1.00004$&$0.74$&$1.5$&$0.75$\\
$4$&$0$  &$1/2$&$-0.39191$&$-0.39194$&$1.00007$&$0.66$&$1.3$&$0.67$\\
$4$&$1/4$&$1/4$&$-0.3856$&$-0.3862$&$1.0015   $&$0.66$&$1.6$&$0.64$\\
$4$&$1/4$&$1/3$&$-0.3887$&$-0.3893$&$1.0016   $&$0.61$&$1.5$&$0.6$\\
$4$&$1/4$&$1/2$&$-0.3962$&$-0.3976$&$1.0036   $&$0.55$&$1.3$&$0.53$\\
$4$&$1/3$&$1/3$&$-0.3918$&$-0.3936$&$1.0045   $&$0.57$&$1.5$&$0.56$\\
$4$&$1/3$&$1/2$&$-0.3993$&$-0.4008$&$1.0037   $&$0.51$&$1.3$&$0.5$\\
$4$&$1/2$&$1/2$&$-0.4068$&$-0.4107$&$1.0096   $&$0.45$&$1.3$&$0.44$\\
\hline
\end{tabular}
%
%
\end{center}
\caption{
Left: the value of $A$ in \eqref{CBA} computed with \eqref{Afinitek} and the results of fitting.
Right: the leading instanton exponent $\omega$ estimated by fitting, which disagree with the leading exponents expected from WKB expansion $2/(1+\eta)$ but rather agree with the conjectural worldsheet instanton exponent $4/(k(1+\xi)(1+\eta))$.
}
\label{omegafit}
\end{table}
\section{Discussion}
\label{discuss}
In this paper we have studied the partition function of a continuous deformation of the $U(N)$ circular quiver superconformal Chern-Simons theories.
The deformation corresponds to the general $R$-charge assignments on the bifundamental matter fields.
Formally the partition function of the deformed theory have similar structures as the superconformal case.
We can use the Fermi gas formalism to compute the large $N$ expansion of the partition function.
Applying this technique to the deformation of the ${\cal N}=4$ theory with the levels characterized by two integers $q$ and $p$ through \eqref{kins} and \eqref{sminimal}, we have achieved to determine the all order perturbative corrections to the partition function in $1/N$, which sum up to an Airy function.
The restriction to ${\cal N}=4$ theories with the special choice of the levels \eqref{sminimal} allows us to completely determine the coefficients $B$ and $A$ appearing in the Airy function, as well as to discover the five series of the non-perturbative effects (membrane instantons).
We find a beautiful decomposition structure of $A$ \eqref{CBA} which is a natural generalization of the similar structure in the undeformed theories observed in \cite{MN1}.
Though there are no clear explanation so far, the rule of decomposition is strongly correlated to the subdivision of the membrane instantons.\footnote{
Similar decomposition structure was observed also in the superconformal theories with affine D-type quiver \cite{MN4} and those with $O(N)$ and $USp(N)$ gauge groups \cite{O}}
We wish to provide some interpretation to this structure in future.

We can also determine the instanton coefficients in the limit $k\rightarrow 0$ while exactly in the other parameters $(q,p,\xi,\eta)$.
We find the singular structure of the coefficients with respect to these parameters, which are again reminiscent of the superconformal case.
The major difference from the superconformal case is the subdivision of the membrane instantons \eqref{instcoefs} by the two {\it continuous} deformation parameters $(\xi,\eta)$ \eqref{xieta}.
Correspondingly, the instanton coefficients diverge for special values of $\xi$ and $\eta$.
The interpretation of these divergences is conceptually different from those appearing in the superconformal case.
In the original superconformal theory $q$ and $p$ are associated to the number of vertices and must be positive integers.
Especially, since they characterize the orbifold structure of $Y_7$ \cite{IK} in the gravity side, there are no dual geometry corresponding to the non-integral $(q,p)$.
Though we can visualize the divergence of the coefficients by continuing these parameters to irrational numbers in the analysis of the matrix model, at the physical values of $(k,q,p)$ there only remain the finite coefficients after the pole cancellation.
In contrast, the $R$-charge assignments $\xi$ and $\eta$ can be chosen to be arbitrary real number under \eqref{xietabound}, and thus the coefficient of the individual instanton can be infinitely large.
This implies that the divergences would be meaningful phenomena in the context of the AdS/CFT correspondence.

On the other hand, our deformation will drastically modify the dual geometry.
Any non-canonical choices of the $R$-charges break the conformal symmetry, which induce in the gravity side non-trivial dependence of the geometry on the radial direction or the holographic RG flow \cite{FP,PTW}.
It will be interesting to reveal the dual geometry to our theory, construct an instanton solution in that background and reveal what occurs near the special values of $(q,p,\xi,\eta)$ where the individual instanton coefficients \eqref{instcoefs} diverge in the field theory side.

We have also analyzed the non-perturbative effects in $1/N$ for finite $k$, which disagree with the membrane instantons.
We have concluded them to be the analog of the worldsheet instantons and have conjectured the exponents as $e^{-\frac{4\ell\mu}{k(q\pm\xi)(p\pm\eta)}}$ $(\ell=1,2,\cdots)$.
Though we have confirmed the exponents for $k=4$ (and also for $k=3,6$ with a few undisplayed data), we could not determine the exact values of their coefficients.
We wish to determine these effects more quantitatively in future.
It would also be interesting to consider similar continuous deformation for the theories with non-circular quiver diagrams \cite{ADF,MN4} or non-unitary gauge groups \cite{MePu,MoSu,O}

\section*{Acknowledgement}
The author is grateful to Louise Anderson, Jun Bourdier, Jan Felix, Masazumi Honda, Sanefumi Moriyama, Tokiro Numasawa, Kazuma Shimizu and Seiji Terashima for valuable discussion.
The authors would also like to appreciate the organizers of the workshop YITP-W-15-12 ``Developments in String Theory and Quantum Field Theory'' held in the Yukawa Institute for Theoretical Physics.
Discussions during the workshop were helpful to complete this work.
The work of the author is partly supported by the JSPS Research Fellowships for Young Scientist.

\appendix
\section{List of ${\cal Z}_{2s}(n)$ in WKB expansion}
\label{apcalZ}
\begin{align}
{\cal Z}_0(n)&=\frac{1}{2\pi}
\B\biggl[\frac{q+\xi}{2}n,\frac{q-\xi}{2}n\biggr]
\B\biggl[\frac{p+\eta}{2}n,\frac{p-\eta}{2}n\biggr],\nonumber \\
{\cal Z}_2(n)&=-\frac{n^2(-1+n^2)(q^2-\xi^2)(p^2-\eta^2)}{384(1+qn)(1+pn)}{\cal Z}_0(n),\nonumber \\
{\cal Z}_4(n)&=\frac{n^2(1-n^2)(q^2-\xi^2)(p^2-\eta^2)}{92160(1+qn)(1+pn)}
\biggl[
\frac{(8q+3n(q^2-\xi^2))(8p+3n(p^2-\eta^2))}{16(3+qn)(3+pn)}(-9+n^2)\nonumber \\
&\quad\quad+\frac{((2q+n(q^2-\xi^2))(2p+n(p^2-\eta^2))}{(2+qn)(2+pn)}(4-n^2)\biggr]
{\cal Z}_0(n),\nonumber \\
{\cal Z}_6(n)&=
\scalebox{0.5}{$\displaystyle
\frac{n^3(1-n^2)(q^2-\xi^2)(p^2-\eta^2)}{11890851840(1+np)(2+np)(3+np)(4+np)(5+np)(1+nq)(2+nq)(3+nq)(4+nq)(5+nq)}
\biggl[
368640 q^4 p + 368640 q p^4 + 1376256 q p n + 1032192 q^3 p n + 276480 q^5 p n
$}\nonumber \\
&\quad \scalebox{0.5}{$\displaystyle
+ 2752512 q^2 p^2 n + 841728 q^4 p^2 n + 1032192 q p^3 n + 442368 q^3 p^3 n + 841728 q^2 p^4 n + 276480 q p^5 n + 3784704 q^2 p n^2 + 884736 q^4 p n^2 + 46080 q^6 p n^2 + 3784704 q p^2 n^2 + 6279168 q^3 p^2 n^2
$}\nonumber \\
&\quad \scalebox{0.5}{$\displaystyle
+ 631296 q^5 p^2 n^2 + 6279168 q^2 p^3 n^2 + 1119744 q^4 p^3 n^2 + 884736 q p^4 n^2 + 1119744 q^3 p^4 n^2 + 631296 q^2 p^5 n^2 + 46080 q p^6 n^2 + 3268608 q^3 p n^3 + 211968 q^5 p n^3 + 9719808 q^2 p^2 n^3
$}\nonumber \\
&\quad \scalebox{0.5}{$\displaystyle
+ 4024320 q^4 p^2 n^3 + 105216 q^6 p^2 n^3 + 3268608 q p^3 n^3 + 9323520 q^3 p^3 n^3 + 646272 q^5 p^3 n^3 + 4024320 q^2 p^4 n^3 + 937536 q^4 p^4 n^3 + 211968 q p^5 n^3 + 646272 q^3 p^5 n^3 + 105216 q^2 p^6 n^3
$}\nonumber \\
&\quad \scalebox{0.5}{$\displaystyle
+ 1228800 q^4 p n^4 + 13824 q^6 p n^4 + 8128512 q^3 p^2 n^4 + 916224 q^5 p^2 n^4 + 8128512 q^2 p^3 n^4 + 5114112 q^4 p^3 n^4 + 98496 q^6 p^3 n^4 + 1228800 q p^4 n^4 + 5114112 q^3 p^4 n^4 + 365904 q^5 p^4 n^4
$}\nonumber \\
&\quad \scalebox{0.5}{$\displaystyle
+ 916224 q^2 p^5 n^4 + 365904 q^4 p^5 n^4 + 13824 q p^6 n^4 + 98496 q^3 p^6 n^4 + 211968 q^5 p n^5 + 2992128 q^4 p^2 n^5 + 64896 q^6 p^2 n^5 + 6678528 q^3 p^3 n^5 + 1103232 q^5 p^3 n^5 + 2992128 q^2 p^4 n^5
$}\nonumber \\
&\quad \scalebox{0.5}{$\displaystyle
+ 2585280 q^4 p^4 n^5 + 47448 q^6 p^4 n^5 + 211968 q p^5 n^5 + 1103232 q^3 p^5 n^5 + 106164 q^5 p^5 n^5 + 64896 q^2 p^6 n^5 + 47448 q^4 p^6 n^5 + 13824 q^6 p n^6 + 507648 q^5 p^2 n^6 + 2426112 q^4 p^3 n^6 + 78336 q^6 p^3 n^6
$}\nonumber \\
&\quad \scalebox{0.5}{$\displaystyle
+ 2426112 q^3 p^4 n^6 + 537600 q^5 p^4 n^6 + 507648 q^2 p^5 n^6 + 537600 q^4 p^5 n^6 + 11574 q^6 p^5 n^6 + 13824 q p^6 n^6 + 78336 q^3 p^6 n^6 + 11574 q^5 p^6 n^6 + 32640 q^6 p^2 n^7 + 407040 q^5 p^3 n^7 + 871680 q^4 p^4 n^7
$}\nonumber \\
&\quad \scalebox{0.5}{$\displaystyle
+ 37872 q^6 p^4 n^7 + 407040 q^3 p^5 n^7 + 109608 q^5 p^5 n^7 + 32640 q^2 p^6 n^7 + 37872 q^4 p^6 n^7 + 1101 q^6 p^6 n^7 + 25920 q^6 p^3 n^8 + 144816 q^5 p^4 n^8 + 144816 q^4 p^5 n^8 + 7668 q^6 p^5 n^8 + 25920 q^3 p^6 n^8
$}\nonumber \\
&\quad \scalebox{0.5}{$\displaystyle
+ 7668 q^5 p^6 n^8 + 9144 q^6 p^4 n^9 + 23844 q^5 p^5 n^9 + 9144 q^4 p^6 n^9 + 534 q^6 p^6 n^9 + 1494 q^6 p^5 n^{10} + 1494 q^5 p^6 n^{10} + 93 q^6 p^6 n^{11} + 3686400 q p^2 \eta^2 + 5160960 q p n \eta^2 + 2211840 q^3 p n \eta^2
$}\nonumber \\
&\quad \scalebox{0.5}{$\displaystyle
+ 8417280 q^2 p^2 n \eta^2 + 1044480 q p^3 n \eta^2 + 10752000 q^2 p n^2 \eta^2 + 1820160 q^4 p n^2 \eta^2 + 847872 q p^2 n^2 \eta^2 + 7769088 q^3 p^2 n^2 \eta^2 + 2384896 q^2 p^3 n^2 \eta^2 - 92160 q p^4 n^2 \eta^2 - 2580480 q p n^3 \eta^2
$}\nonumber \\
&\quad \scalebox{0.5}{$\displaystyle
+ 6620160 q^3 p n^3 \eta^2 + 397440 q^5 p n^3 \eta^2 + 353280 q^2 p^2 n^3 \eta^2 + 3178752 q^4 p^2 n^3 \eta^2 - 374784 q p^3 n^3 \eta^2 + 1937664 q^3 p^3 n^3 \eta^2 - 210432 q^2 p^4 n^3 \eta^2 - 6236160 q^2 p n^4 \eta^2 + 1470720 q^4 p n^4 \eta^2
$}\nonumber \\
&\quad \scalebox{0.5}{$\displaystyle
+ 20160 q^6 p n^4 \eta^2 - 1732608 q p^2 n^4 \eta^2 - 1665792 q^3 p^2 n^4 \eta^2 + 511488 q^5 p^2 n^4 \eta^2 - 1453568 q^2 p^3 n^4 \eta^2 + 649504 q^4 p^3 n^4 \eta^2 - 27648 q p^4 n^4 \eta^2 - 196992 q^3 p^4 n^4 \eta^2 - 5099520 q^3 p n^5 \eta^2
$}\nonumber \\
&\quad \scalebox{0.5}{$\displaystyle
+ 101760 q^5 p n^5 \eta^2 - 4162560 q^2 p^2 n^5 \eta^2 - 1437888 q^4 p^2 n^5 \eta^2 + 21312 q^6 p^2 n^5 \eta^2 - 374784 q p^3 n^5 \eta^2 - 1744896 q^3 p^3 n^5 \eta^2 + 71880 q^5 p^3 n^5 \eta^2 - 129792 q^2 p^4 n^5 \eta^2 - 94896 q^4 p^4 n^5 \eta^2
$}\nonumber \\
&\quad \scalebox{0.5}{$\displaystyle
- 1862400 q^4 p n^6 \eta^2 - 3375360 q^3 p^2 n^6 \eta^2 - 369504 q^5 p^2 n^6 \eta^2 - 894464 q^2 p^3 n^6 \eta^2 - 863168 q^4 p^3 n^6 \eta^2 - 644 q^6 p^3 n^6 \eta^2 - 27648 q p^4 n^6 \eta^2 - 156672 q^3 p^4 n^6 \eta^2 - 23148 q^5 p^4 n^6 \eta^2
$}\nonumber \\
&\quad \scalebox{0.5}{$\displaystyle
- 314880 q^5 p n^7 \eta^2 - 1220160 q^4 p^2 n^7 \eta^2 - 29088 q^6 p^2 n^7 \eta^2 - 718080 q^3 p^3 n^7 \eta^2 - 179280 q^5 p^3 n^7 \eta^2 - 65280 q^2 p^4 n^7 \eta^2 - 75744 q^4 p^4 n^7 \eta^2 - 2202 q^6 p^4 n^7 \eta^2 - 20160 q^6 p n^8 \eta^2
$}\nonumber \\
&\quad \scalebox{0.5}{$\displaystyle
- 204192 q^5 p^2 n^8 \eta^2 - 256352 q^4 p^3 n^8 \eta^2 - 12824 q^6 p^3 n^8 \eta^2 - 51840 q^3 p^4 n^8 \eta^2 - 15336 q^5 p^4 n^8 \eta^2 - 12960 q^6 p^2 n^9 \eta^2 - 42360 q^5 p^3 n^9 \eta^2 - 18288 q^4 p^4 n^9 \eta^2 - 1068 q^6 p^4 n^9 \eta^2 - 2660 q^6 p^3 n^{10} \eta^2
$}\nonumber \\
&\quad \scalebox{0.5}{$\displaystyle
- 2988 q^5 p^4 n^{10} \eta^2 - 186 q^6 p^4 n^{11} \eta^2 + 1843200 q \eta^4 + 4208640 q^2 n \eta^4 - 1320960 q p n \eta^4 - 2027520 q n^2 \eta^4 + 2833920 q^3 n^2 \eta^4 - 3016192 q^2 p n^2 \eta^4 + 46080 q p^2 n^2 \eta^4 - 4500480 q^2 n^3 \eta^4
$}\nonumber \\
&\quad \scalebox{0.5}{$\displaystyle
+ 749760 q^4 n^3 \eta^4 + 162816 q p n^3 \eta^4 - 2583936 q^3 p n^3 \eta^4 + 105216 q^2 p^2 n^3 \eta^4 + 552960 q n^4 \eta^4 - 3294720 q^3 n^4 \eta^4 + 85680 q^5 n^4 \eta^4 + 537344 q^2 p n^4 \eta^4 - 1015408 q^4 p n^4 \eta^4 + 13824 q p^2 n^4 \eta^4
$}\nonumber \\
&\quad \scalebox{0.5}{$\displaystyle
+ 98496 q^3 p^2 n^4 \eta^4 + 1305600 q^2 n^5 \eta^4 - 1044480 q^4 n^5 \eta^4 + 4200 q^6 n^5 \eta^4 + 162816 q p n^5 \eta^4 + 641664 q^3 p n^5 \eta^4 - 178044 q^5 p n^5 \eta^4 + 64896 q^2 p^2 n^5 \eta^4 + 47448 q^4 p^2 n^5 \eta^4 + 1059840 q^3 n^6 \eta^4
$}\nonumber \\
&\quad \scalebox{0.5}{$\displaystyle
- 151200 q^5 n^6 \eta^4 + 386816 q^2 p n^6 \eta^4 + 325568 q^4 p n^6 \eta^4 - 10930 q^6 p n^6 \eta^4 + 13824 q p^2 n^6 \eta^4 + 78336 q^3 p^2 n^6 \eta^4 + 11574 q^5 p^2 n^6 \eta^4 + 386880 q^4 n^7 \eta^4 - 8400 q^6 n^7 \eta^4 + 311040 q^3 p n^7 \eta^4
$}\nonumber \\
&\quad \scalebox{0.5}{$\displaystyle
+ 69672 q^5 p n^7 \eta^4 + 32640 q^2 p^2 n^7 \eta^4 + 37872 q^4 p^2 n^7 \eta^4 + 1101 q^6 p^2 n^7 \eta^4 + 65520 q^5 n^8 \eta^4 + 111536 q^4 p n^8 \eta^4 + 5156 q^6 p n^8 \eta^4 + 25920 q^3 p^2 n^8 \eta^4 + 7668 q^5 p^2 n^8 \eta^4 + 4200 q^6 n^9 \eta^4
$}\nonumber \\
&\quad \scalebox{0.5}{$\displaystyle
+ 18516 q^5 p n^9 \eta^4 + 9144 q^4 p^2 n^9 \eta^4 + 534 q^6 p^2 n^9 \eta^4 + 1166 q^6 p n^{10} \eta^4 + 1494 q^5 p^2 n^{10} \eta^4 + 93 q^6 p^2 n^{11} \eta^4 + 3686400 q^2 p \xi^2 + 5160960 q p n \xi^2 + 1044480 q^3 p n \xi^2 + 8417280 q^2 p^2 n \xi^2
$}\nonumber \\
&\quad \scalebox{0.5}{$\displaystyle
+ 2211840 q p^3 n \xi^2 + 847872 q^2 p n^2 \xi^2 - 92160 q^4 p n^2 \xi^2 + 10752000 q p^2 n^2 \xi^2 + 2384896 q^3 p^2 n^2 \xi^2 + 7769088 q^2 p^3 n^2 \xi^2 + 1820160 q p^4 n^2 \xi^2 - 2580480 q p n^3 \xi^2 - 374784 q^3 p n^3 \xi^2
$}\nonumber \\
&\quad \scalebox{0.5}{$\displaystyle
+ 353280 q^2 p^2 n^3 \xi^2 - 210432 q^4 p^2 n^3 \xi^2 + 6620160 q p^3 n^3 \xi^2 + 1937664 q^3 p^3 n^3 \xi^2 + 3178752 q^2 p^4 n^3 \xi^2 + 397440 q p^5 n^3 \xi^2 - 1732608 q^2 p n^4 \xi^2 - 27648 q^4 p n^4 \xi^2 - 6236160 q p^2 n^4 \xi^2
$}\nonumber \\
&\quad \scalebox{0.5}{$\displaystyle
- 1453568 q^3 p^2 n^4 \xi^2 - 1665792 q^2 p^3 n^4 \xi^2 - 196992 q^4 p^3 n^4 \xi^2 + 1470720 q p^4 n^4 \xi^2 + 649504 q^3 p^4 n^4 \xi^2 + 511488 q^2 p^5 n^4 \xi^2 + 20160 q p^6 n^4 \xi^2 - 374784 q^3 p n^5 \xi^2 - 4162560 q^2 p^2 n^5 \xi^2
$}\nonumber \\
&\quad \scalebox{0.5}{$\displaystyle
- 129792 q^4 p^2 n^5 \xi^2 - 5099520 q p^3 n^5 \xi^2 - 1744896 q^3 p^3 n^5 \xi^2 - 1437888 q^2 p^4 n^5 \xi^2 - 94896 q^4 p^4 n^5 \xi^2 + 101760 q p^5 n^5 \xi^2 + 71880 q^3 p^5 n^5 \xi^2 + 21312 q^2 p^6 n^5 \xi^2 - 27648 q^4 p n^6 \xi^2
$}\nonumber \\
&\quad \scalebox{0.5}{$\displaystyle
- 894464 q^3 p^2 n^6 \xi^2 - 3375360 q^2 p^3 n^6 \xi^2 - 156672 q^4 p^3 n^6 \xi^2 - 1862400 q p^4 n^6 \xi^2 - 863168 q^3 p^4 n^6 \xi^2 - 369504 q^2 p^5 n^6 \xi^2 - 23148 q^4 p^5 n^6 \xi^2 - 644 q^3 p^6 n^6 \xi^2 - 65280 q^4 p^2 n^7 \xi^2
$}\nonumber \\
&\quad \scalebox{0.5}{$\displaystyle
- 718080 q^3 p^3 n^7 \xi^2 - 1220160 q^2 p^4 n^7 \xi^2 - 75744 q^4 p^4 n^7 \xi^2 - 314880 q p^5 n^7 \xi^2 - 179280 q^3 p^5 n^7 \xi^2 - 29088 q^2 p^6 n^7 \xi^2 - 2202 q^4 p^6 n^7 \xi^2 - 51840 q^4 p^3 n^8 \xi^2 - 256352 q^3 p^4 n^8 \xi^2
$}\nonumber \\
&\quad \scalebox{0.5}{$\displaystyle
- 204192 q^2 p^5 n^8 \xi^2 - 15336 q^4 p^5 n^8 \xi^2 - 20160 q p^6 n^8 \xi^2 - 12824 q^3 p^6 n^8 \xi^2 - 18288 q^4 p^4 n^9 \xi^2 - 42360 q^3 p^5 n^9 \xi^2 - 12960 q^2 p^6 n^9 \xi^2 - 1068 q^4 p^6 n^9 \xi^2 - 2988 q^4 p^5 n^{10} \xi^2 - 2660 q^3 p^6 n^{10} \xi^2
$}\nonumber \\
&\quad \scalebox{0.5}{$\displaystyle
- 186 q^4 p^6 n^{11} \xi^2 + 11059200 q p n \eta^2 \xi^2 + 1059840 q^2 p n^2 \eta^2 \xi^2 + 1059840 q p^2 n^2 \eta^2 \xi^2 - 12057600 q p n^3 \eta^2 \xi^2 - 1017600 q^3 p n^3 \eta^2 \xi^2 - 3608832 q^2 p^2 n^3 \eta^2 \xi^2 - 1017600 q p^3 n^3 \eta^2 \xi^2
$}\nonumber \\
&\quad \scalebox{0.5}{$\displaystyle
- 5149440 q^2 p n^4 \eta^2 \xi^2 - 40320 q^4 p n^4 \eta^2 \xi^2 - 5149440 q p^2 n^4 \eta^2 \xi^2 - 1491200 q^3 p^2 n^4 \eta^2 \xi^2 - 1491200 q^2 p^3 n^4 \eta^2 \xi^2 - 40320 q p^4 n^4 \eta^2 \xi^2 + 3532800 q p n^5 \eta^2 \xi^2 - 399360 q^3 p n^5 \eta^2 \xi^2
$}\nonumber \\
&\quad \scalebox{0.5}{$\displaystyle
- 806592 q^2 p^2 n^5 \eta^2 \xi^2 - 42624 q^4 p^2 n^5 \eta^2 \xi^2 - 399360 q p^3 n^5 \eta^2 \xi^2 - 398256 q^3 p^3 n^5 \eta^2 \xi^2 - 42624 q^2 p^4 n^5 \eta^2 \xi^2 + 2430720 q^2 p n^6 \eta^2 \xi^2 + 2430720 q p^2 n^6 \eta^2 \xi^2 + 462976 q^3 p^2 n^6 \eta^2 \xi^2
$}\nonumber \\
&\quad \scalebox{0.5}{$\displaystyle
+ 462976 q^2 p^3 n^6 \eta^2 \xi^2 + 1288 q^4 p^3 n^6 \eta^2 \xi^2 + 1288 q^3 p^4 n^6 \eta^2 \xi^2 + 541440 q^3 p n^7 \eta^2 \xi^2 + 1636800 q^2 p^2 n^7 \eta^2 \xi^2 + 58176 q^4 p^2 n^7 \eta^2 \xi^2 + 541440 q p^3 n^7 \eta^2 \xi^2 + 272736 q^3 p^3 n^7 \eta^2 \xi^2
$}\nonumber \\
&\quad \scalebox{0.5}{$\displaystyle
+ 58176 q^2 p^4 n^7 \eta^2 \xi^2 + 4404 q^4 p^4 n^7 \eta^2 \xi^2 + 40320 q^4 p n^8 \eta^2 \xi^2 + 355456 q^3 p^2 n^8 \eta^2 \xi^2 + 355456 q^2 p^3 n^8 \eta^2 \xi^2 + 25648 q^4 p^3 n^8 \eta^2 \xi^2 + 40320 q p^4 n^8 \eta^2 \xi^2 + 25648 q^3 p^4 n^8 \eta^2 \xi^2 + 25920 q^4 p^2 n^9 \eta^2 \xi^2
$}\nonumber \\
&\quad \scalebox{0.5}{$\displaystyle
+ 74832 q^3 p^3 n^9 \eta^2 \xi^2 + 25920 q^2 p^4 n^9 \eta^2 \xi^2 + 2136 q^4 p^4 n^9 \eta^2 \xi^2 + 5320 q^4 p^3 n^{10} \eta^2 \xi^2 + 5320 q^3 p^4 n^{10} \eta^2 \xi^2 + 372 q^4 p^4 n^{11} \eta^2 \xi^2 - 4723200 q n^2 \eta^4 \xi^2 - 2058240 q^2 n^3 \eta^4 \xi^2 + 620160 q p n^3 \eta^4 \xi^2
$}\nonumber \\
&\quad \scalebox{0.5}{$\displaystyle
+ 3340800 q n^4 \eta^4 \xi^2 - 191520 q^3 n^4 \eta^4 \xi^2 + 979712 q^2 p n^4 \eta^4 \xi^2 + 20160 q p^2 n^4 \eta^4 \xi^2 + 1852800 q^2 n^5 \eta^4 \xi^2 - 8400 q^4 n^5 \eta^4 \xi^2 + 297600 q p n^5 \eta^4 \xi^2 + 326376 q^3 p n^5 \eta^4 \xi^2 + 21312 q^2 p^2 n^5 \eta^4 \xi^2
$}\nonumber \\
&\quad \scalebox{0.5}{$\displaystyle
- 691200 q n^6 \eta^4 \xi^2 + 302400 q^3 n^6 \eta^4 \xi^2 - 93472 q^2 p n^6 \eta^4 \xi^2 + 21860 q^4 p n^6 \eta^4 \xi^2 - 644 q^3 p^2 n^6 \eta^4 \xi^2 - 485760 q^2 n^7 \eta^4 \xi^2 + 16800 q^4 n^7 \eta^4 \xi^2 - 226560 q p n^7 \eta^4 \xi^2 - 93456 q^3 p n^7 \eta^4 \xi^2
$}\nonumber \\
&\quad \scalebox{0.5}{$\displaystyle
- 29088 q^2 p^2 n^7 \eta^4 \xi^2 - 2202 q^4 p^2 n^7 \eta^4 \xi^2 - 110880 q^3 n^8 \eta^4 \xi^2 - 151264 q^2 p n^8 \eta^4 \xi^2 - 10312 q^4 p n^8 \eta^4 \xi^2 - 20160 q p^2 n^8 \eta^4 \xi^2 - 12824 q^3 p^2 n^8 \eta^4 \xi^2 - 8400 q^4 n^9 \eta^4 \xi^2 - 32472 q^3 p n^9 \eta^4 \xi^2
$}\nonumber \\
&\quad \scalebox{0.5}{$\displaystyle
- 12960 q^2 p^2 n^9 \eta^4 \xi^2 - 1068 q^4 p^2 n^9 \eta^4 \xi^2 - 2332 q^4 p n^{10} \eta^4 \xi^2 - 2660 q^3 p^2 n^{10} \eta^4 \xi^2 - 186 q^4 p^2 n^{11} \eta^4 \xi^2 + 1843200 p \xi^4 - 1320960 q p n \xi^4 + 4208640 p^2 n \xi^4 - 2027520 p n^2 \xi^4 + 46080 q^2 p n^2 \xi^4
$}\nonumber \\
&\quad \scalebox{0.5}{$\displaystyle
- 3016192 q p^2 n^2 \xi^4 + 2833920 p^3 n^2 \xi^4 + 162816 q p n^3 \xi^4 - 4500480 p^2 n^3 \xi^4 + 105216 q^2 p^2 n^3 \xi^4 - 2583936 q p^3 n^3 \xi^4 + 749760 p^4 n^3 \xi^4 + 552960 p n^4 \xi^4 + 13824 q^2 p n^4 \xi^4 + 537344 q p^2 n^4 \xi^4
$}\nonumber \\
&\quad \scalebox{0.5}{$\displaystyle
- 3294720 p^3 n^4 \xi^4 + 98496 q^2 p^3 n^4 \xi^4 - 1015408 q p^4 n^4 \xi^4 + 85680 p^5 n^4 \xi^4 + 162816 q p n^5 \xi^4 + 1305600 p^2 n^5 \xi^4 + 64896 q^2 p^2 n^5 \xi^4 + 641664 q p^3 n^5 \xi^4 - 1044480 p^4 n^5 \xi^4 + 47448 q^2 p^4 n^5 \xi^4
$}\nonumber \\
&\quad \scalebox{0.5}{$\displaystyle
- 178044 q p^5 n^5 \xi^4 + 4200 p^6 n^5 \xi^4 + 13824 q^2 p n^6 \xi^4 + 386816 q p^2 n^6 \xi^4 + 1059840 p^3 n^6 \xi^4 + 78336 q^2 p^3 n^6 \xi^4 + 325568 q p^4 n^6 \xi^4 - 151200 p^5 n^6 \xi^4 + 11574 q^2 p^5 n^6 \xi^4 - 10930 q p^6 n^6 \xi^4
$}\nonumber \\
&\quad \scalebox{0.5}{$\displaystyle
+ 32640 q^2 p^2 n^7 \xi^4 + 311040 q p^3 n^7 \xi^4 + 386880 p^4 n^7 \xi^4 + 37872 q^2 p^4 n^7 \xi^4 + 69672 q p^5 n^7 \xi^4 - 8400 p^6 n^7 \xi^4 + 1101 q^2 p^6 n^7 \xi^4 + 25920 q^2 p^3 n^8 \xi^4 + 111536 q p^4 n^8 \xi^4 + 65520 p^5 n^8 \xi^4
$}\nonumber \\
&\quad \scalebox{0.5}{$\displaystyle
+ 7668 q^2 p^5 n^8 \xi^4 + 5156 q p^6 n^8 \xi^4 + 9144 q^2 p^4 n^9 \xi^4 + 18516 q p^5 n^9 \xi^4 + 4200 p^6 n^9 \xi^4 + 534 q^2 p^6 n^9 \xi^4 + 1494 q^2 p^5 n^{10} \xi^4 + 1166 q p^6 n^{10} \xi^4 + 93 q^2 p^6 n^{11} \xi^4 - 4723200 p n^2 \eta^2 \xi^4
$}\nonumber \\
&\quad \scalebox{0.5}{$\displaystyle
+ 620160 q p n^3 \eta^2 \xi^4 - 2058240 p^2 n^3 \eta^2 \xi^4 + 3340800 p n^4 \eta^2 \xi^4 + 20160 q^2 p n^4 \eta^2 \xi^4 + 979712 q p^2 n^4 \eta^2 \xi^4 - 191520 p^3 n^4 \eta^2 \xi^4 + 297600 q p n^5 \eta^2 \xi^4 + 1852800 p^2 n^5 \eta^2 \xi^4 + 21312 q^2 p^2 n^5 \eta^2 \xi^4
$}\nonumber \\
&\quad \scalebox{0.5}{$\displaystyle
+ 326376 q p^3 n^5 \eta^2 \xi^4 - 8400 p^4 n^5 \eta^2 \xi^4 - 691200 p n^6 \eta^2 \xi^4 - 93472 q p^2 n^6 \eta^2 \xi^4 + 302400 p^3 n^6 \eta^2 \xi^4 - 644 q^2 p^3 n^6 \eta^2 \xi^4 + 21860 q p^4 n^6 \eta^2 \xi^4 - 226560 q p n^7 \eta^2 \xi^4 - 485760 p^2 n^7 \eta^2 \xi^4
$}\nonumber \\
&\quad \scalebox{0.5}{$\displaystyle
- 29088 q^2 p^2 n^7 \eta^2 \xi^4 - 93456 q p^3 n^7 \eta^2 \xi^4 + 16800 p^4 n^7 \eta^2 \xi^4 - 2202 q^2 p^4 n^7 \eta^2 \xi^4 - 20160 q^2 p n^8 \eta^2 \xi^4 - 151264 q p^2 n^8 \eta^2 \xi^4 - 110880 p^3 n^8 \eta^2 \xi^4 - 12824 q^2 p^3 n^8 \eta^2 \xi^4 - 10312 q p^4 n^8 \eta^2 \xi^4
$}\nonumber \\
&\quad \scalebox{0.5}{$\displaystyle
- 12960 q^2 p^2 n^9 \eta^2 \xi^4 - 32472 q p^3 n^9 \eta^2 \xi^4 - 8400 p^4 n^9 \eta^2 \xi^4 - 1068 q^2 p^4 n^9 \eta^2 \xi^4 - 2660 q^2 p^3 n^{10} \eta^2 \xi^4 - 2332 q p^4 n^{10} \eta^2 \xi^4 - 186 q^2 p^4 n^{11} \eta^2 \xi^4 + 1339200 n^3 \eta^4 \xi^4 + 105840 q n^4 \eta^4 \xi^4
$}\nonumber \\
&\quad \scalebox{0.5}{$\displaystyle
+ 105840 p n^4 \eta^4 \xi^4 - 777600 n^5 \eta^4 \xi^4 + 4200 q^2 n^5 \eta^4 \xi^4 - 148332 q p n^5 \eta^4 \xi^4 + 4200 p^2 n^5 \eta^4 \xi^4 - 151200 q n^6 \eta^4 \xi^4 - 151200 p n^6 \eta^4 \xi^4 - 10930 q^2 p n^6 \eta^4 \xi^4 - 10930 q p^2 n^6 \eta^4 \xi^4 + 129600 n^7 \eta^4 \xi^4
$}\nonumber \\
&\quad \scalebox{0.5}{$\displaystyle
- 8400 q^2 n^7 \eta^4 \xi^4 + 23784 q p n^7 \eta^4 \xi^4 - 8400 p^2 n^7 \eta^4 \xi^4 + 1101 q^2 p^2 n^7 \eta^4 \xi^4 + 45360 q n^8 \eta^4 \xi^4 + 45360 p n^8 \eta^4 \xi^4 + 5156 q^2 p n^8 \eta^4 \xi^4 + 5156 q p^2 n^8 \eta^4 \xi^4 + 4200 q^2 n^9 \eta^4 \xi^4 + 13956 q p n^9 \eta^4 \xi^4
$}\nonumber \\
&\quad \scalebox{0.5}{$\displaystyle
+ 4200 p^2 n^9 \eta^4 \xi^4 + 534 q^2 p^2 n^9 \eta^4 \xi^4 + 1166 q^2 p n^{10} \eta^4 \xi^4 + 1166 q p^2 n^{10} \eta^4 \xi^4 + 93 q^2 p^2 n^{11} \eta^4 \xi^4
\biggr]{\cal Z}_0(n)
$}.
\label{calZs}
\end{align}
Here $\B$ is the Euler beta function
\begin{align}
\B(x,y)=\frac{\Gamma(x)\Gamma(y)}{\Gamma(x+y)}.
\label{Beta}
\end{align}

\section{List of exact values $Z^{(\xi;\eta)}_4(N)$}
\label{evlist}
\begin{align}
Z^{(0;1/4)}_4(1)&=\frac{\sqrt{2}}{8}\sin\frac{\pi}{8},\quad
Z^{(0;1/4)}_4(2)=\frac{-4+3\sqrt{2}}{512},\nonumber \\
Z^{(0;1/4)}_4(3)&=\frac{16(2+\sqrt{2})+(1-13\sqrt{2})\pi}{2048\pi}\sin\frac{\pi}{8},\quad
Z^{(0;1/4)}_4(4)=\frac{-128(2+\sqrt{2})+(9+92\sqrt{2})\pi}{262144\pi},
\label{Zexactsk41}
\end{align}
\begin{align}
Z^{(0;1/3)}_4(1)&=\frac{\sqrt{3}}{24},\quad
Z^{(0;1/3)}_4(2)=\frac{1}{1728},\quad
Z^{(0;1/3)}_4(3)=\frac{-216+(216-85\sqrt{3})\pi}{20736\pi},\nonumber \\
Z^{(0;1/3)}_4(4)&=\frac{4320\sqrt{3}+(-137-1296\sqrt{3})\pi}{5971968\pi},\quad
Z^{(0;1/3)}_4(5)=\frac{-10584+(25056-12521\sqrt{3})\pi}{35831808\pi},
\label{Zexactsk42}
\end{align}
\begin{align}
Z^{(0;1/2)}_4(1)&=\frac{\sqrt{2}}{16},\quad
Z^{(0;1/2)}_4(2)=\frac{4-\pi}{256\pi},\quad
Z^{(0;1/2)}_4(3)=\frac{-8\sqrt{2}+12\sqrt{2}\pi-3\sqrt{2}\pi^2}{4096\pi^2},\nonumber \\
Z^{(0;1/2)}_4(4)&=\frac{-304-120\pi+69\pi^2}{393216\pi^2},\nonumber \\
Z^{(0;1/2)}_4(5)&=\frac{-192\sqrt{2}-960\sqrt{2}\pi-3424\sqrt{2}\pi^2-1832\sqrt{2}\pi^3+963\sqrt{2}\pi^4}{18874368\pi^4},\nonumber \\
Z^{(0;1/2)}_4(6)&=\frac{-3840-20160\pi+42640\pi^2+36236\pi^3-15165\pi^4}{1509949440\pi^4},\nonumber \\
Z^{(0;1/2)}_4(7)&=\frac{23040\sqrt{2}-241920\sqrt{2}\pi+3019200\sqrt{2}\pi^2-7891200\sqrt{2}\pi^3-1138088\sqrt{2}\pi^4}{1087163596800\pi^6}\nonumber \\
&\quad +\frac{6964164\sqrt{2}\pi^5-1877175\sqrt{2}\pi^6}{1087163596800\pi^6},\nonumber \\
Z^{(0;1/2)}_4(8)&=\frac{-3870720+16441600\pi+85612800\pi^2-107341600\pi^3-115970736\pi^4}{54116587929600\pi^5}\nonumber \\
&\quad+\frac{44873325\pi^5}{54116587929600\pi^5},
\label{Zexactsk43}
\end{align}
\begin{align}
Z^{(1/4;1/4)}_4(1)&=\frac{2-\sqrt{2}}{8},\quad
Z^{(1/4;1/4)}_4(2)=\frac{1-2\sqrt{2}+2(1+\sqrt{2})\sin\frac{\pi}{8}}{32},\nonumber \\
Z^{(1/4;1/4)}_4(3)&=\frac{-(4+3\sqrt{2})\sin\frac{\pi}{8}+(4-(5+2\sqrt{2})\sin\frac{\pi}{8})\pi}{128\pi},
\label{Zexactsk44}
\end{align}
\begin{align}
Z^{(1/4;1/3)}_4(1)&=\frac{\sqrt{6}}{12}\sin\frac{\pi}{8},\quad
Z^{(1/4;1/3)}_4(2)=\frac{-1-\sqrt{2}+\sqrt{6}}{48},\nonumber \\
Z^{(1/4;1/3)}_4(3)&=\frac{72+36\sqrt{2}+(18+81\sqrt{2}-68\sqrt{3}-22\sqrt{6})\pi}{1728\pi}\sin\frac{\pi}{8},\nonumber \\
Z^{(1/4;1/3)}_4(4)&=\frac{24\sqrt{6}+\Bigl(228-77\sqrt{2}+84\sqrt{3}-120\sqrt{6}+3\sqrt{6(443-180\sqrt{6})}\Bigr)\pi}{27648\pi},
\label{Zexactsk45}
\end{align}
\begin{align}
Z^{(1/4;1/2)}_4(1)&=\frac{1}{4}\sin\frac{\pi}{8},\quad
Z^{(1/4;1/2)}_4(2)=\frac{3-2\sqrt{2}}{128},\nonumber \\
Z^{(1/4;1/2)}_4(3)&=\frac{-32-16\sqrt{2}+(16+\sqrt{2})\pi}{2048\pi}\sin\frac{\pi}{8},\quad
Z^{(1/4;1/2)}_4(4)=\frac{-64+(43-16\sqrt{2})\pi}{131072\pi},\nonumber \\
Z^{(1/4;1/2)}_4(5)&=\frac{768+768\sqrt{2}+(3424+3712\sqrt{2})\pi+(-2079-615\sqrt{2})\pi^2}{3145728\pi^2}\sin\frac{\pi}{8},
\label{Zexactsk46}
\end{align}
\begin{align}
Z^{(1/3;1/3)}_4(1)=\frac{1}{12},\quad
Z^{(1/3;1/3)}_4(2)=\frac{-3+4\sqrt{3}\sin\frac{\pi}{9}+4\sin\frac{\pi}{18}}{72},
\label{Zexactsk47}
\end{align}
\begin{align}
Z^{(1/3;1/2)}_4(1)&=\frac{\sqrt{6}}{24},\quad
Z^{(1/3;1/2)}_4(2)=\frac{-3+2\sqrt{3}}{288},\quad
Z^{(1/3;1/2)}_4(3)=\frac{144\sqrt{2}+(-96\sqrt{2}+29\sqrt{6})\pi}{13824\pi},\nonumber \\
Z^{(1/3;1/2)}_4(4)&=\frac{-64\sqrt{3}+(63-16\sqrt{3})\pi}{221184\pi},\nonumber \\
Z^{(1/3;1/2)}_4(5)&=\frac{-1344\sqrt{2}+384\sqrt{6}+(-1632\sqrt{2}+1067\sqrt{6})\pi}{5308416\pi},\nonumber \\
Z^{(1/3;1/2)}_4(6)&=\frac{311040+(1009152+159840\sqrt{3})\pi-1141425\pi^2+404470\sqrt{3}\pi^2}{2866544640\pi^2},
\label{Zexactsk48}
\end{align}
\begin{align}
Z^{(1/2;1/2)}_4(1)&=\frac{1}{8},\quad
Z^{(1/2;1/2)}_4(2)=\frac{-2+\pi}{128\pi},\quad
Z^{(1/2;1/2)}_4(3)=\frac{2-10\pi+3\pi^2}{1024\pi^2},\nonumber \\
Z^{(1/2;1/2)}_4(4)&=\frac{70+6\pi-9\pi^2}{49152\pi^2},\quad
Z^{(1/2;1/2)}_4(5)=\frac{3+66\pi+406\pi^2+175\pi^3-99\pi^4}{589824\pi^4}.
\label{Zexactsk49}
\end{align}

\end{document}